\RequirePackage{fancyvrb}

\documentclass[onefignum,onetabnum]{siamart190516}

\usepackage{amsfonts}
\usepackage{algorithmic}
\usepackage{mathtools}
\usepackage{subcaption}
\usepackage{booktabs}
\usepackage[inline]{enumitem}

\VerbatimFootnotes
\graphicspath{{img/}}
\ifpdf
  \DeclareGraphicsExtensions{.eps,.pdf,.png,.jpg}
\else
  \DeclareGraphicsExtensions{.eps}
\fi
\newlist{parenum}{enumerate}{1}
\setlist[parenum]{leftmargin=0cm,itemindent=2\parindent,noitemsep,label=(\arabic*)}
\newlist{inlinenum}{enumerate*}{1}
\setlist[inlinenum]{label=(\arabic*)}

\newcommand\str{\textnormal{\texttt{Stratum}}}
\newcommand\ROOT{\textnormal{\texttt{root}}}
\newcommand\LEAF{\textnormal{\texttt{leaf}}}

\headers{Fully Parallel Mesh I/O using PETSc DMPlex}{V. Hapla et al.}

\title{Fully Parallel Mesh I/O using PETSc DMPlex with an Application to Waveform Modeling\thanks{Revised version submitted to SISC on September 14, 2020.}}

\author{
Vaclav Hapla\thanks{ETH Zurich, Switzerland (\email{vaclav.hapla@erdw.ethz.ch}, \email{andreas.fichtner@erdw.ethz.ch}).}
\and Matthew G. Knepley\thanks{University at Buffalo, NY (\email{knepley@buffalo.edu}).}
\and Michael Afanasiev\footnotemark[2]
\and Christian Boehm\footnotemark[2]
\and Martin van Driel\footnotemark[2]
\and Lion Krischer\footnotemark[2]
\and Andreas Fichtner\footnotemark[2]
}

\usepackage{amsopn}

\begin{document}

\maketitle

\begin{abstract}
Large-scale PDE simulations using high-order finite-element methods on unstructured meshes are an indispensable tool in science and engineering.
The widely used open-source PETSc library offers an efficient representation of generic unstructured meshes within its DMPlex module.
This paper details our recent implementation of parallel mesh reading and topological interpolation (computation of edges and faces from a cell-vertex mesh) into DMPlex.
We apply these developments to seismic wave propagation scenarios on Mars as an example application.
The principal motivation is to overcome single-node memory limits and reach mesh sizes which were impossible before.
Moreover, we demonstrate that scalability of I/O and topological interpolation goes beyond 12'000 cores, and memory-imposed limits on mesh size vanish.
\end{abstract}



\section{Introduction}
Finite-element methods (FEM) \cite{Larson_book_2013,Hughes_book_2000,Zienkiewicz_book_2013}
are widely used in science and engineering to simulate complex physical systems.
Many applications require FEM dicretization with high polynomial order on large unstructured meshes requiring distributed memory computer architectures \cite{Afanasiev_2019,beriot_efficient_2016,rudi_extreme-scale_2015,karna_thetis_2018,mfem_paper}.
Our main motivating application here is a spectral-element method implementation \cite{Afanasiev_2019} where several requirements arise:

\begin{parenum}
\item \label{req1} The distributed memory mesh representation should be {\em fully connected}, i.e. explicitly include mesh entities of all topological codimensions\footnote{
  In 3D, all vertices, edges, faces (= facets) and regions (= cells).
  In 2D, all vertices, edges (= facets) and faces (= cells).
},
so that adjacency queries involving any two codimensions are equally efficient.
However, mesh data files typically store only vertices and cells to avoid redundant storage, disk operations, and to conform to widely used formats.
Edges and faces can be computed in runtime using a process which will be called here {\em topological interpolation}.

\item \label{req2} Meshing tools typically create a single data file,
but loading the whole mesh onto one computational node and distributing onto all remaining nodes becomes a bottleneck for a sufficiently large mesh due to memory constraints\footnote{
  On our testing platform, Piz Daint operated by the Swiss National Supercomputing Centre, the mesh size limit was approximately 16 million hexahedra elements.
}.
Instead, we need to load different portions of the mesh file directly onto target processors and maintain a distributed mesh representation right from beginning.
For sake of flexibility and optimal use of current computational resources,
we do not want to depend on an a priori partitioning information stored in the mesh file,
nor even storing partitions in separate files.

\item \label{req3} We need to be able to redistribute the loaded, naively distributed mesh to reduce halo communication in the simulation phase.

\item \label{req4} The mesh file format should be widely supported and offer good interoperability among tools such as visualization or mesh conversion software.
It must be naturally suitable for parallel I/O as per \ref{req2}.
\end{parenum}

PETSc DMPlex (\cref{sec:dmplex}) is a flexible mesh implementation which meets these criteria.
A mesh is represented as a graph whose vertices represent mesh entities, and its edges represent their incidence relations.
DMPlex is agnostic to the mesh shape, dimensionality and mesh entity types, and can represent any number of codimensions.
Topological interpolation had been implemented, so \ref{req1} was readily addressed.
Parallel mesh partitioner interfaces had also been already available in DMPlex, hence there was nothing to solve for \ref{req3} neither.
Moreover, the requirement \ref{req4} is easy to meet since PETSc already possesses interfaces for many mesh formats and HDF5;
we have chosen XDMF on top of HDF5 because it is widely supported and HDF5 is well known to support efficient parallel I/O natively.
The missing piece, however, was to address \ref{req2}, which is the main topic of this paper.
Along with parallel loading, {\em parallel} topological interpolation needed to be implemented as well.

We demonstrate the efficiency and potential of our new DMPlex functionality on a real-world application in the context of the full-waveform modeling.
The spectral-element method on unstructured conforming hexahedral meshes has become the de-facto standard for global-scale simulations of seismic waves \cite{Afanasiev_2019,Ferroni_2017,Fichtner_book,Peter_2011}.
We apply it to simulate full 3D high-frequency wave propagation on Mars, based on data from the NASA InSight mission \cite{Banerdt2020}.
This consists in solving a coupled system of the elastic and acoustic wave equations.
To accurately model these data in the desired frequency band, large scale simulations are required.
Moreover, the ability of the mesh framework to represent all codimensions is crucial here.
In the presented simulation, more than 100 million 4-th order hexahedral mesh elements have been used,
which is substantially beyond the original limit of 16 million elements.

The manuscript is organized as follows.
First, we briefly introduce related efforts by other teams.
Then we describe abstractions for mesh data management of unstructured meshes in high-order finite-element discretizations using PETSc DMPlex.
Further, we explain our new strategies for the parallel simulation startup (mesh reading, topological interpolation and redistribution) on distributed memory HPC architectures.
We continue with a brief introduction of the spectral-element method and its implementation which uses DMPlex.
Next, waveform modeling benchmarks demonstrate scalable performance of the parallel startup for up to 256 million hexahedral mesh elements,
running on up to 1024 Cray XC50 nodes of Piz Daint.
Finally, the mentioned Mars seismic wave propagation simulation is presented as an application.

\section{Related work}
Similar efforts within flexible distributed mesh infrastructures have been made by other authors.

The mesh representation in {\em DOLFIN}, a part of the FEniCS project~\cite{fenics,fenics-paper},
is inspired by DMPlex and uses similar concepts~\cite{logg_efficient_2009,LoggWells2010a}\footnote{
  These publications refer to the DMPlex predecessor {\em Sieve}~\cite{knepley_mesh_2009}.
}.
This implementation supports parallel I/O as well~\cite{richardson_parallel_slides}.

{\em MOAB}~\cite{moab,moab_users}\footnote{
  Note PETSc includes interface to MOAB (\verb|DMMOAB|).
} is a component for representing and evaluating structured or unstructured meshes.
The MOAB API is ``designed to be simple yet powerful''.
MOAB is ``optimized for runtime and storage efficiency, based on access to mesh in chunks rather than through individual topological entities,
while also versatile enough to support individual entity access''.
Entities are addressed through integer handles rather than pointers, to allow the underlying entity to change without changing the handle.
The handle encodes the entity type in four bits.
Hence the set of supported types is fixed.
In the standard MOAB database, only vertices and cells are explicitly represented, and non-vertex entity to cell adjacency queries are expensive.
This limitation can be mitigated by the optional Array-based Half-Facet (AHF) representation.
The associated maps are created during the first adjacency query, which makes it substantially more expensive than the subsequent queries.
AHF supports only a limited class of meshes, e.g. only uniform cell types.
MOAB uses its own mesh file format.
Although it is based on HDF5, a prior partitioning (assigning entities to per-process sets) is needed to load a mesh file in parallel~\cite{moab_users}.
The partitioning is done by a separate serial program which stores the information into the file~\cite{moab_users}.
Zoltan \cite{ZoltanHomePage,ZoltanUsersGuideV3} is employed within the partitioning utility.

{\em PUMI} (Parallel Unstructured Mesh Infrastructure)~\cite{pumi_users,pumi_2012,pumi_2009} is an unstructured, distributed mesh data management system
``capable of handling general non-manifold models and effectively supporting automated adaptive analysis''.
Stress is put on flexibility of the mesh representation with respect to generating adjacencies between different codimensions.
However, this product is again able to read in parallel only with a pre-partitioned mesh, one file per partition,
and currently supports only its own specific mesh file format~\cite{pumi_users}.

{\em STK Mesh} is a part of Sierra Toolkit (STK)~\cite{sierra_toolkit,sierra_toolkit_users}.
STK modules are intended to provide infrastructure that assists the development of computational engineering software such as finite-element analysis applications.
STK includes modules for unstructured-mesh data structures, reading/writing mesh files, geometric proximity search, and various utilities.
The primary file format is Exodus~II~\cite{exodus}.
Fully connected mesh representation is supported.
Mesh entities are internally organized into ``buckets''.
The entities in a bucket all have the same codimension and shape, and they are all members of the same mesh parts.
Buckets correspond to contiguously-allocated blocks of memory in the associated field-data values.
The design is well suited for run-time mesh modifications.
Ghost cell layers of any thickness are supported in distributed meshes.
Run-time parallel repartitioning using ParMETIS~\cite{parmetis,karypis_fast_1998,karypis_parallel_1998} is available.
STK Mesh does not currently provide parallel mesh file I/O~\cite{sierra_toolkit_users}.

{\em MSTK}~\cite{mstk,mstk_users,garimella_mesh_2002} is a flexible mesh framework,
offering several mesh representation with a common interface.
One of them is the fully connected representation F1 which is claimed to be the most mature.
Reduced representations are available, though, with the missing entities created on-the-fly transparently if needed.
These ``volatile'' entities are cached for certain time to accelerate possible reuse.
Thanks to this design, MSTK can always be called as if the mesh was fully connected.
MSTK is designed in an object oriented manner.
MSTK has ``preliminary support for representation of distributed surface and volume meshes'';
parallel modification of meshes is not yet supported~\cite{mstk_users}.
MSTK supports its own mesh file format as well as GMV~\cite{gmv} and Exodus~II~\cite{exodus}.
Serial partitioning can be done using METIS~\cite{metis,karypis_fast_1998} interface.
Neither parallel repartitioning nor parallel I/O support is currently implemented~\cite{mstk_users}.

In this paper, we do not deal with hierarchical mesh representations such as octree data structures \cite{flaherty_adaptive_1997,schneiders_octree-based_2000}.
They are advantageous for certain data access patterns, especially in the context of Adaptive Mesh Refinement (AMR) \cite{flaherty_adaptive_1997,burstedde_extreme-scale_2010},
at the expense of generality;
DMPlex supports, for instance, arbitrary mesh partitions and extraction of arbitrary subsets of cells (or facets) as submeshes,
features which are typically missing from hierarchical meshing frameworks \cite{Isaac_2015}.
Note that PETSc offers the DMForest wrapper of p4est \cite{p4est,p4est_recursive,rudi_extreme-scale_2015}, and conversion between DMPlex and DMForest \cite{Isaac_2015}.

\section{DMPlex}\label{sec:dmplex}
PETSc~\cite{petsc-web-page,petsc-user-ref,petsc-efficient,brown_extensibility_2015} is a well-known library
for numerical methods, used by the scientific and engineering computing communities.
It provides parallel data management, structured and unstructured meshes,
linear and nonlinear algebraic solvers and preconditioners, time integrators, optimization algorithms and others.
Many of these methods (such as geometric multigrid and domain decomposition solvers) can take advantage
of the geometric/topological setting of a discretized problem, i.e., mesh information.

DMPlex is a PETSc module for generic unstructured mesh storage and operations.
It decouples user applications from the implementation details of common mesh-related utility tasks,
such as file I/O and mesh partitioning.
It represents the mesh topology in a flexible way,
providing topological connectivity of mesh entities at all codimensions (vertices, edges, faces and cells),
crucial for high-order finite-element method (FEM) simulations,
and provides a wide range of common mesh management functionalities.

In the rest of the paper, we will sometimes refer to specific PETSc functions so that an interested reader
can look up working implementations in the PETSc reference manual\footnote{
  \url{https://www.mcs.anl.gov/petsc/petsc-dev/docs/manualpages/singleindex.html}
}.

PETSc's interface for serving mesh data to numerical algorithms is the {\tt DM} object. PETSc has several {\tt DM} implementations.
The native implementations of structured grids ({\tt DMDA}), staggered grids {\tt DMStag}, and unstructured meshes ({\tt DMPlex}) have the most
complete coverage of the {\tt DM} API, and are developed most actively.
Besides these two, PETSc has several {\tt DM} implementations that wrap external libraries,
such as {\tt DMMOAB} for MOAB~\cite{moab} and {\tt DMFOREST} for p4est~\cite{p4est}.
Here we will focus on {\tt DMPlex} which proved to be most relevant for the discussed waveform modeling applications.

{\tt DMPlex} encapsulates the topology of unstructured grids and provides a wide range of common mesh management functionalities
to application programmers~\cite{rathgeber_firedrake_2016,lange_efficient_2016,lange_flexible_2015,barral_anisotropic_2016,dalcin_petiga_2016,Afanasiev_2019}.
It provides a domain topology abstraction that decouples user applications from the implementation details of common mesh-related utility tasks, such as file I/O, mesh generation, partitioning, and parallel repartitioning.
It aims to increase extensibility and interoperability between scientific applications through librarization~\cite{petsc-efficient,brown_extensibility_2015}.

\subsection{Mesh representation and basic operations}\label{sec:dmplex_basic}
\begin{figure}
    \centering
    \begin{subfigure}{0.35\columnwidth}
        \centering
        \includegraphics[width=\linewidth]{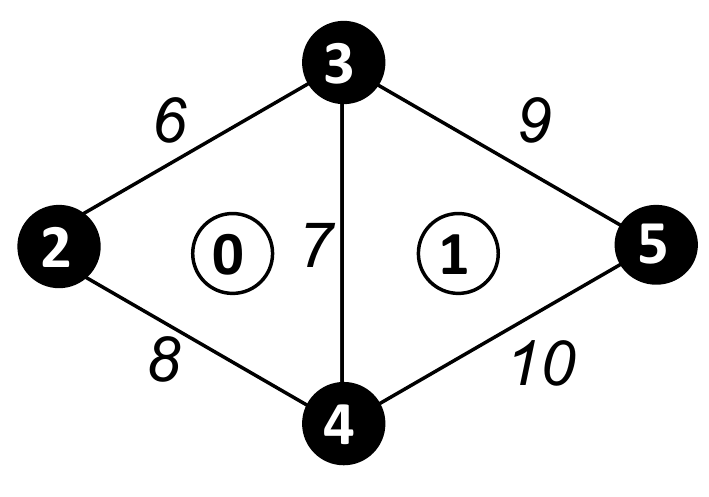}
        \caption{original 2D mesh}
        \label{fig:seq_mesh_int0}
    \end{subfigure}
    \hfill
    \begin{subfigure}{0.35\columnwidth}
        \centering
        \includegraphics[width=\linewidth]{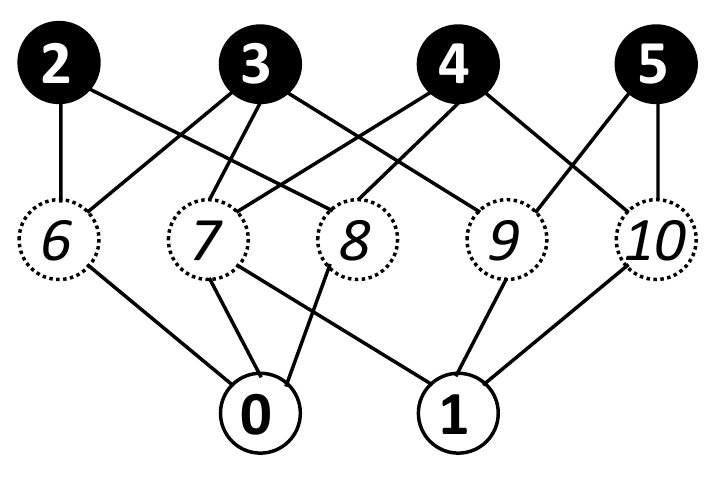}
        \caption{plex representation of \subref{fig:seq_mesh_int0}, having 3 strata}
        \label{fig:seq_mesh_int_plex0}
    \end{subfigure}
    \hfill
    \begin{subfigure}{0.28\columnwidth}
        \centering
        \includegraphics[height=7.8em]{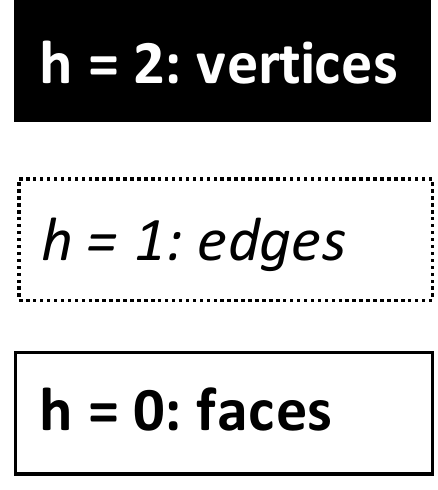}
        \caption{coloring of $\str(h)$, $h=0,1,2$, throughout the paper}
        \label{fig:strata_legend}
    \end{subfigure}
    \\
    \begin{subfigure}{0.49\columnwidth}
        \centering
        \includegraphics[width=.8\linewidth]{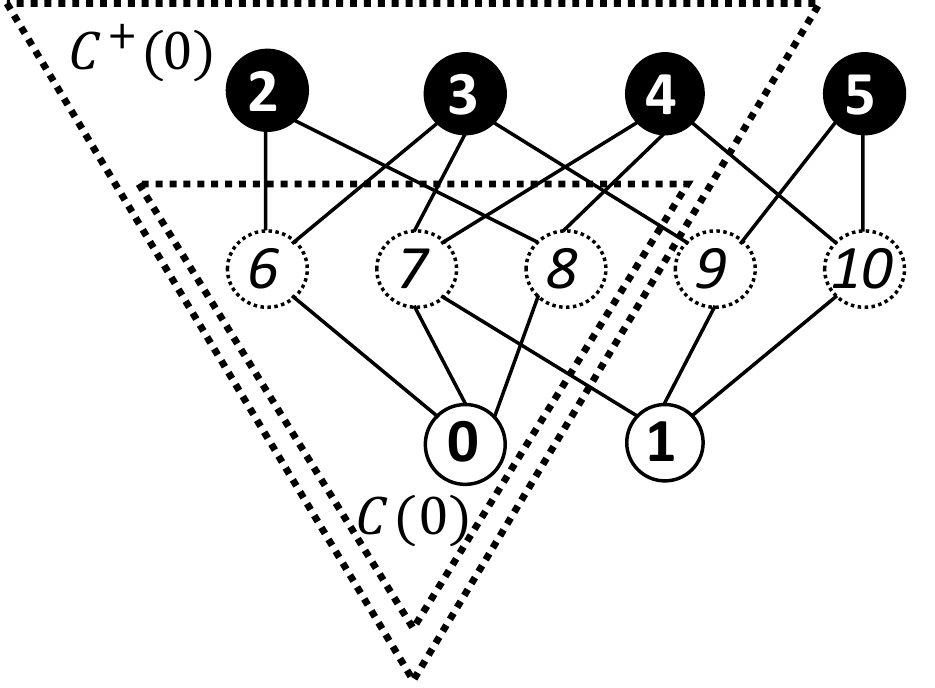}
        \caption{cone $C(0)$,\\its transitive closure $C^+(0)$ }
        \label{fig:cone}
    \end{subfigure}
    \hfill
    \begin{subfigure}{0.49\columnwidth}
        \centering
        \includegraphics[width=.8\linewidth]{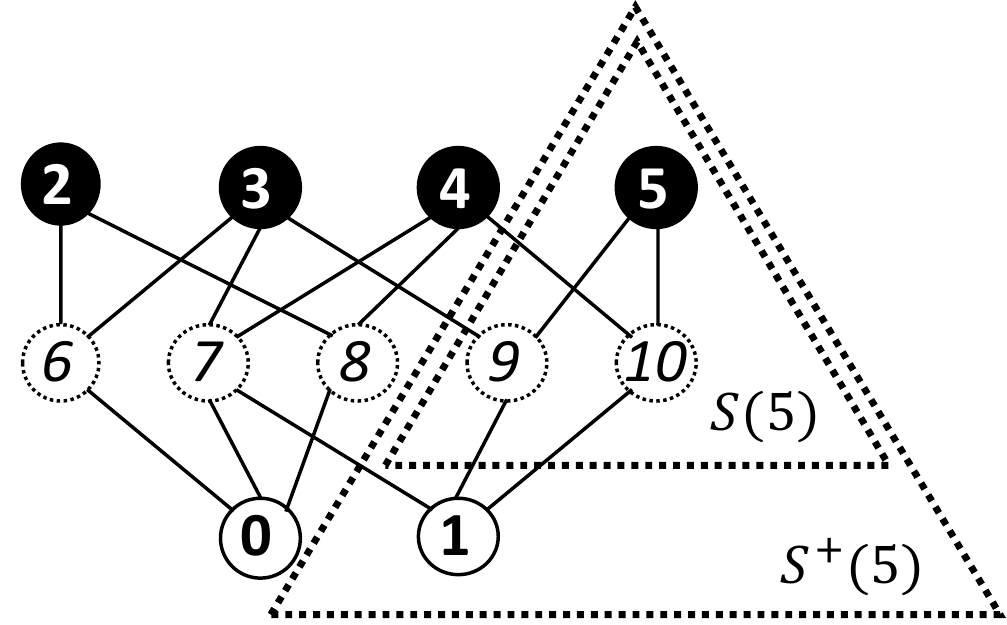}
        \caption{support $S(5)$,\\its transitive closure $S^+(5)$}
        \label{fig:support}
    \end{subfigure}
    \caption{DMPlex mesh representation and basic relations.
      The DAG stratum height $h$ corresponds to topological codimension.
      A 1D mesh would have 2 strata: $h=0$ edges, $h=1$ vertices.
      A 3D mesh would have 4 strata: $h=0$ regions, $h=1$ faces, $h=2$ edges, $h=3$ vertices.
      Note we can also refer to the $h=1$ entities commonly, in a dimension-independent way, as {\em facets},
      and $h=0$ as {\em cells}.
      The entities at $h=d$, where $d$ is the topological dimension, are always vertices.
    }
    \label{fig:strata}
\end{figure}

{\tt DMPlex} uses an abstract representation of the unstructured mesh topology,
where topological entities (vertices, edges, facets, cells) form vertices of a directed acyclic graph
(DAG)~\cite{logg_efficient_2009,knepley_mesh_2009},
also known as Hasse Diagram.
Edges in this graph represent incidence of entities of two subsequent represented codimensions,
e.g. incidence of a facet with a cell\footnote{
  The difference of subsequent codimensions can be more than 1.
  For example, the DAG can directly connect a vertex with a face, if it does not store edges.
}.
Let us call this representation a {\em plex} and refer to the topological mesh entities as DAG points or just {\em points}.
The plex is constructed of clearly defined independent sets ({\em strata}) that correspond to different topological codimensions, see \cref{fig:seq_mesh_int_plex0}.

The strata can be assigned a contiguous integer numbering $h$ called {\em height}, $h = 0, 1, \ldots, H \leq d$,
where $H$ is the total number of strata or total height\footnote{
  $H$ can be computed using depth-first graph search.
},
and $d$ is the embedding topological dimension of the mesh.
Without loss of generality, we can choose the numbering to follow the bottom-up order in the graph drawing.
Let us denote $\str(h)$ a set of points in the same stratum at height $h$.

The plex which represents all mesh entities of all codimensions is called {\em fully connected}\footnote{
  Corresponding to the full representation F1 in MSTK~\cite{mstk_users,garimella_mesh_2002}.
  A fully connected plex represents all vertices, edges, faces (= facets) and regions (= cells) in the 3D case.
  In 2D, all vertices, edges (= facets) and faces (= cells).
  The fully connected plex can be computed from an initial reduced plex (representing vertices and cells only)
  by means of topological interpolation described in \cref{sec:interpolation}.
}.
For such plex, $H = d$,
and $\str(h)$ then, mnemonically, contains all and only those points which represent entities of codimension $h$.
See \cref{fig:seq_mesh_int_plex0,fig:strata_legend} for illustration.
The reverse (up-bottom) strata numbering can be called {\em depth} and it corresponds to the topological dimension.

All plex points share a
single consecutive numbering, emphasizing that each point is treated equally
no matter its shape or dimension, and allowing us to store the graph connectivity
in a single array where dimensional layers are defined as consecutively numbered
subranges. The directional connectivity of the plex is defined by the covering relation
called {\em cone}, denoted here as $C(p)$,
yielding a set of all points directly connected to $p$ in the next higher stratum,
i.e. representing incident entities of the next higher codimension\footnote{
  ``Next higher'' codimension automatically means ``higher by 1'' only for a fully connected plex, otherwise the difference can be more than 1.
}.
The transitive closure of the cone relation shall be denoted by $C^+(p)$.
Both $C(p)$ and $C^+(p)$ are illustrated in \cref{fig:cone}.
The dual relation called {\em support} is denoted $S(p)$,
and it gives points directly connected to $p$ from the next lower stratum,
i.e. representing incident entities of the next lower codimension\footnote{
  ``Next lower'' codimension automatically means ``lower by 1'' only for a fully connected plex, otherwise the difference can be more than 1.
}.
Together with its transitive closure $S^+(p)$ it is shown in \cref{fig:support}.

In addition to the abstract topology data, PETSc provides two utility objects
to describe the parallel data layout: a {\tt PetscSection} object maps the graph-based topology
information to discretized solution data through an offset mapping,
and a star forest object\footnote{
  {\tt PetscSF}, see \cref{sec:sf}.
} holds a one-sided description of shared data in parallel. These data layout mappings
allow {\tt DMPlex} to manage distributed solution data by automating the preallocation
of distributed vector and matrix data structures and performing halo data exchanges.
Moreover, by storing grid topology alongside discretized solution data, {\tt DMPlex} is able
to provide the mappings required for sophisticated preconditioning algorithms, such
as geometric multigrid methods~\cite{BruneKnepleyScott2013,farrell_augmented_2019} and multiblock, or ``fieldsplit'' preconditioning
for multiphysics problems~\cite{bkmms2012}.

\subsection{Topological interpolation}\label{sec:interpolation}
For high order methods we are interested in, we need the fully connected mesh representation,
as introduced in \cref{sec:dmplex_basic}, i.e., representing all mesh entities of all codimensions.
However, usual mesh generation algorithms or mesh file readers result in a mesh representation with cells and vertices only\footnote{
  The reduced representation R1 in MSTK~\cite{mstk_users,garimella_mesh_2002}.
},
while edges and faces need to be either handled implicitly \cite{celes_efficient_2005},
or computed explicitly at runtime using {\em topological interpolation} which can be implemented as described further.
While the former is more appealing with respect to the memory efficiency, it brings implementation complexity,
even more substantial for distributed meshes\footnote{
  At the same time, with a scalable distributed mesh implementation, it is easier to overcome the memory limits, which relativizes the gains of the reduced representation.
}.

\begin{figure}
    \centering
    \begin{subfigure}{0.49\columnwidth}
        \centering
        \includegraphics[width=.7\linewidth]{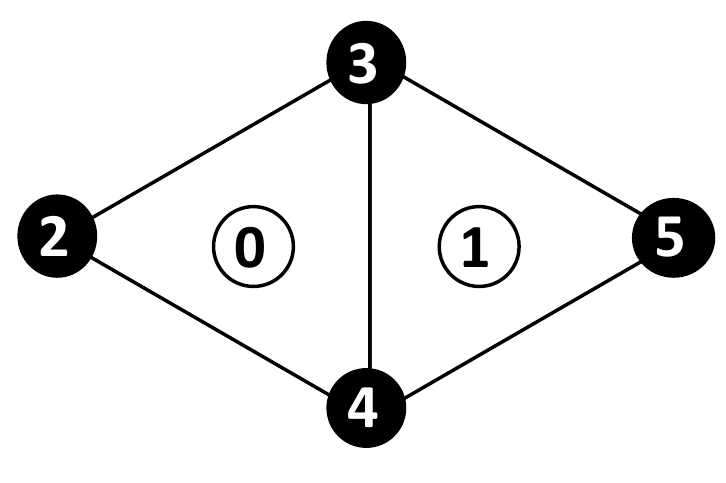}
        \caption{original mesh}
        \label{fig:seq_mesh_orig}
    \end{subfigure}
    \hfill
    \begin{subfigure}{0.49\columnwidth}
        \centering
        \includegraphics[width=.7\linewidth]{seq_mesh_int}
        \caption{mesh \subref{fig:seq_mesh_orig} interpolated}
        \label{fig:seq_mesh_int}
    \end{subfigure}
    \\
    \begin{subfigure}{0.49\columnwidth}
        \centering
        \includegraphics[width=.7\linewidth]{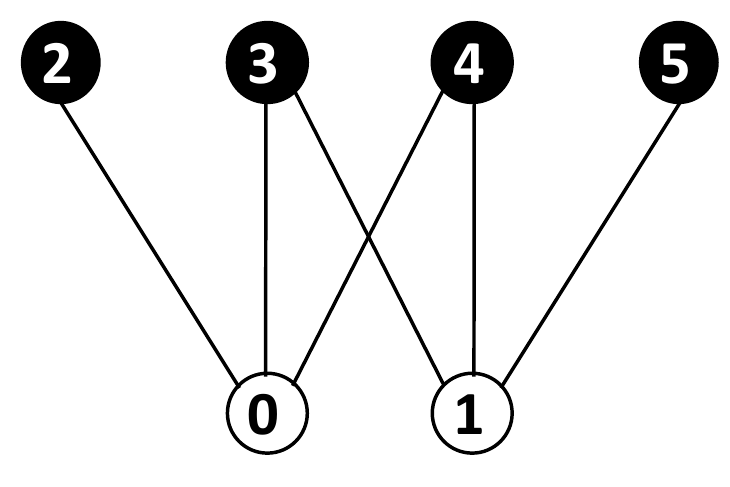}
        \caption{plex representation of \subref{fig:seq_mesh_orig}}
        \label{fig:seq_mesh_plex}
    \end{subfigure}
    \hfill
    \begin{subfigure}{0.49\columnwidth}
        \centering
        \includegraphics[width=.7\linewidth]{seq_mesh_int_plex}
        \caption{plex representation of \subref{fig:seq_mesh_int}}
        \label{fig:seq_mesh_int_plex}
    \end{subfigure}
    \caption{Sequential topological interpolation: original mesh and interpolated mesh in classical and plex representation.}
    \label{fig:seq_mesh}
\end{figure}

Topological interpolation is the process of constructing intermediate levels of the ranked poset describing a mesh,
given information at bracketing levels. For example, if we receive triangles and their covering vertices, as
in~\cref{fig:seq_mesh}, interpolation will construct edges. The first algorithm for interpolation on the Hasse
diagram was published in~\cite{logg_efficient_2009}, but this version is only appropriate for simplices, ignores orientation of the
mesh entities, and did not give a complexity bound.

The topological interpolation procedure selects a given point stratum as cells, for which it will construct facets\footnote{
  So it is applied once in 2D to compute faces, twice in 3D to compute edges and faces.
}.
Briefly, it iterates over the cells, generates oriented facets of each cell from its original cone,
and attaches them to the cell with the correct relative orientation\footnote{
  Detailed in \cref{sec:orient}.
}.
Let us now describe this procedure more in detail.

An initial iteration over cells serves for preallocation.
It takes each cell, computes vertex tuples corresponding to facets of the cell\footnote{
  Using the \verb|DMPlexGetRawFaces_Internal()| helper function.
  It takes as inputs the cell polytope type and a list of vertices forming the cell.
  It returns the cell's facets, namely the polytope type and vertex tuple of each facet.
  For the list of polytope type currently supported,
  see \url{https://www.mcs.anl.gov/petsc/petsc-dev/docs/manualpages/DM/DMPolytopeType.html}.
},
and enters each facet vertex tuple as a hash key into a hash table.
If the key is not yet present in the hash table,
a consecutive DAG point number is assigned to this facet and inserted to the hash table as the value under this key.
Once finished, the hash table size gives two important pieces of information:
\begin{inlinenum}
  \item the number of new facets to be inserted, so that we know the allocation size of the new plex;
  \item a mapping of face vertex tuples to DAG points\footnote{
      if two facets of two different cells have the same vertices (modulo order),
      they are considered the same topological entity and get mapped to the same DAG point.
    }.
\end{inlinenum}

A second iteration over cells inserts the facets into the new plex.
It is identical to the old plex, except it has a new face stratum,
and the cone sizes of cells need to be calculated anew\footnote{
  For example, hexahedra have 8 vertices but 6 facets, so the cone size changes from 8 to 6.
}.
We repeat the face extraction loop above.
We take each cell and extract vertex tuples representing its facets again\footnote{
  Using \verb|DMPlexGetRawFaces_Internal()| again.
}.
We lookup the facet in the table and get its DAG point number so that we can update the cone information.
We insert this facet into the cone of the cell.
Further, if the facet itself has not yet a cone specified, we enter the vertex tuple as it is as this facet's cone.
If instead the facet's cone has been already set, we set the correct relative orientation of the facet with respect to the cell,
computed from comparing the vertex tuple order with the cone order already stored.

The complexity to interpolate a given stratum is in $\mathcal{O}(N_C N_F N_T)$,
where $N_C$ is the number of cells,
$N_F$ is the maximum number of faces per cell,
and $N_T$ is the number of used face types.
However, $N_F$ and $N_T$ are obviously constant, so $\mathcal{O}(N_C N_F N_T)$ = $\mathcal{O}(N_C)$.
The $\mathcal{O}$-complexity remains the same even if we sum all strata because the size of (number of points in) stratum $h=k+1$
is a certain multiple of the size of stratum $h=k$ for any feasible height $k$.
We can conclude that the complexity is linear with respect to the mesh size.

\subsection{Orienting edges and faces}\label{sec:orient}
Let us focus on two particular entities in the interpolated mesh in \cref{fig:seq_mesh_int}: cell 0, and edge $7 \in C(0)$.
Their cones are $C(0) = \{6,7,8\}$ and $C(7) = \{3,4\}$, respectively.
We have so far spoken about \emph{which} points form $C(p)$ for given $p$.
However, since our plex DAG ultimately represents a mesh topology, the order of cone points is also important.
The cone order of a plex point translates to the orientation of the respective topological entity with respect to other entities.
Hence, we will call the cone order {\em orientation} of $p$.
It is needed, for instance, to have a well-defined direction of an outer normal, or to assign field values correctly during simulation.
Hence, $C(0)$ is rather a tuple, $C(0) = (6,7,8)$. This is how the array implementing $C(0)$ is stored.
Let us from now denote by $C(p,c)$ the $c$-th point in this tuple, $c=0,1,\ldots,n-1$, where $n = \text{size}(C(p))$.

\begin{figure}
    \centering
    \begin{subfigure}{0.49\columnwidth}
        \centering
        \includegraphics[width=.7\linewidth]{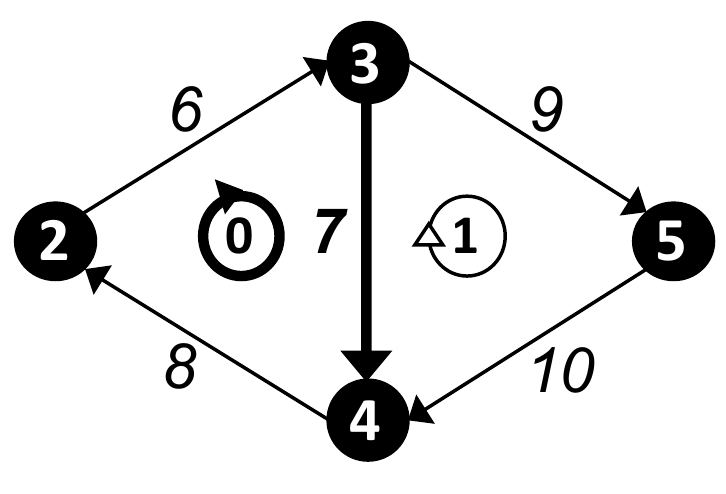}
        \caption{edge 7 in cone of face 0,\\$C(0,1) = 7$}
        \label{fig:ornt_0}
    \end{subfigure}
    \hfill
    \begin{subfigure}{0.49\columnwidth}
        \centering
        \includegraphics[width=.7\linewidth]{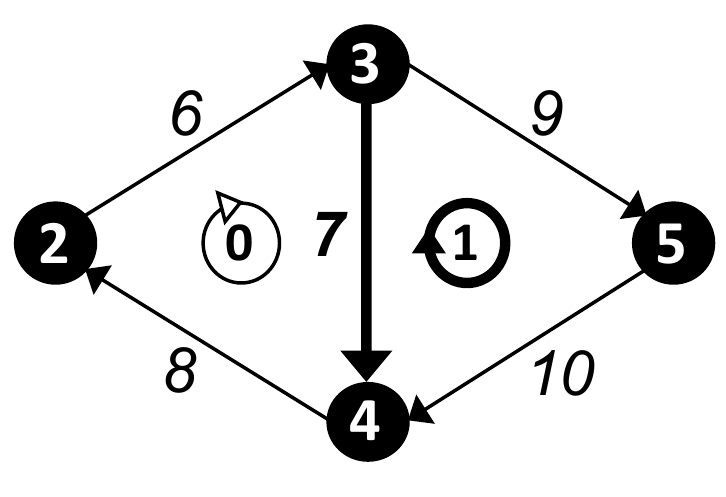}
        \caption{edge 7 in cone of face 1,\\$C(1,0) = 7$}
        \label{fig:ornt_1}
    \end{subfigure}

    \begin{subfigure}{0.49\columnwidth}
        \centering
        \includegraphics[width=.7\linewidth]{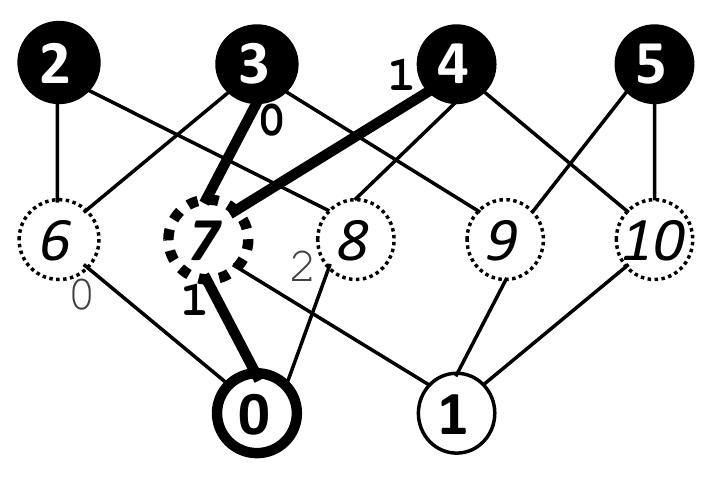}
        \caption{
          the starting point of edge 7 within face 0 is 3;
          $C(C(0,1), S(0,1)) = C(7, 0) = 3,$\\
          $O(0,1) = 0$
        }
        \label{fig:ornt_0_plex}
    \end{subfigure}
    \hfill
    \begin{subfigure}{0.49\columnwidth}
        \centering
        \includegraphics[width=.7\linewidth]{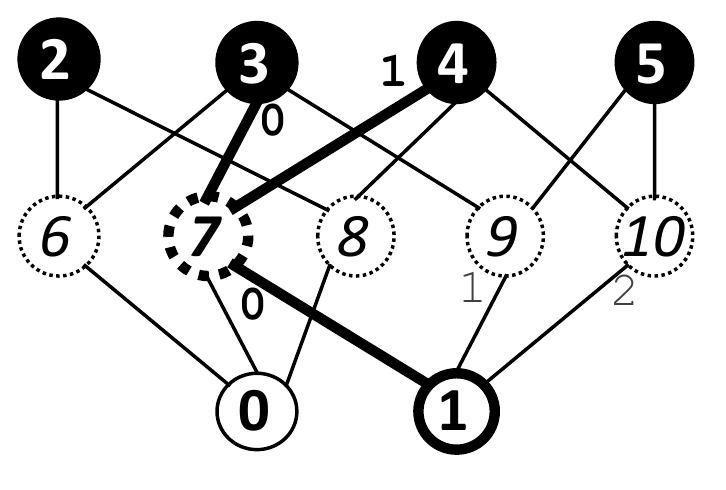}
        \caption{
          the starting point of edge 7 within face 1 is 4;
          $C(C(1,0), S(1,0)) = C(7, 1) = 4,$\\
          $O(1,0) = -2$
        }
        \label{fig:ornt_1_plex}
    \end{subfigure}
    \caption{Mesh from \cref{fig:seq_mesh_int,fig:seq_mesh_int_plex} with cone points order and orientation. We focus here on the edge 7 within the cones of faces 0 and 1.}
    \label{fig:ornt}
\end{figure}

For consistency (e.g. correct evaluation of attached fields), the orientation of points should be in line with the orientation of their supporting points.
This does not concern the lowermost and uppermost stratum; cells have no supporting points, and vertices have no orientation.
However, for the intermediate strata (i.e. edges and faces computed by interpolation), we can get a conflicting situation as depicted in \cref{fig:ornt}.
Edge 7 is oriented against the orientation of cell 1 but if we flipped it, it would be oriented against the orientation of cell 0.
Thus we need a mechanism to allow for this.

In general, suppose point $p \in \str(0)$, its cone point $q = C(p,c) \in C(p)\,\subset\,\str(1)$, and $C(q)\,\subset\,\str(2)$.
For example, in \cref{fig:ornt_1,fig:ornt_1_plex}, $p = 1$, $c = 0$, $q = 7$, $C(q) = (3,4)$.
To compensate the given orientation of $q$ given by the order of $C(q)$, a {\em relative} orientation of $q$ with respect to $p$ needs to be defined.
This information must be attached to the edge ($p$, $q$) in the DAG because it varies for different choices of $p$ even for the same $q$.

The relative orientation can be described by
\begin{inlinenum}
\item the starting point $S(p,c) \geq 0$ in 0-based local numbering with respect to $q$, and
\item the direction $D(p,c) \in \{-1,1\}$ (reverse/forward).
\end{inlinenum}
These two can be represented by a single signed integer $O(p,c)$: $S(p,c)$ by its magnitude, and $D(p,c)$ by its sign.
Since the sign is undefined for 0, the negative values are shifted by -1.
To summarize, the relative orientation shall be defined as
\begin{equation}
O(p,c) =
\begin{cases}
  S(p,c), & D(p,c) = 1,\\
  -S(p,c) - 1, & D(p,c) = -1,
\end{cases}
\end{equation}
and the other way around,
\begin{align}
D(p,c) &=
\begin{cases}
  \mathrlap{1,}\hphantom{-O(p,c) - 1,} & O(p,c) \geq 0,\\
  - 1, & O(p,c) < 0,
\end{cases}\\
S(p,c) &=
\begin{cases}
  O(p,c), & O(p,c) \geq 0,\\
  -O(p,c) - 1, & O(p,c) < 0.
\end{cases}
\end{align}
Note that for $C(p,c)$ being an edge, flipping the orientation of the edge implies changing the starting point, so $O(p,c) \in \{0, -2\}$ only.
We can also define the whole tuple of relative cone point orientations for point $p$,
\begin{equation}
O(p) = \left(O(p,0),\ldots,O(p,n-1)\right),
\end{equation}
where $n = \text{size}(C(p))$. This $O(p)$ is attached to every plex point $p$ in the same way as $C(p)$. We note that
$O(p)$ is just a numbering for the elements of the dihedral group for a given face type.

\section{Parallel simulation startup}\label{sec:startup}
Let us call {\em startup phase} all steps necessary to load the mesh from disk storage and prepare it for use in the simulation time loop.
It consists of the following steps:
\begin{inlinenum}
    \item raw data loading,
    \item plex construction,
    \item topological interpolation,
    \item distribution.
\end{inlinenum}

\begin{figure}
    \includegraphics[width=\columnwidth]{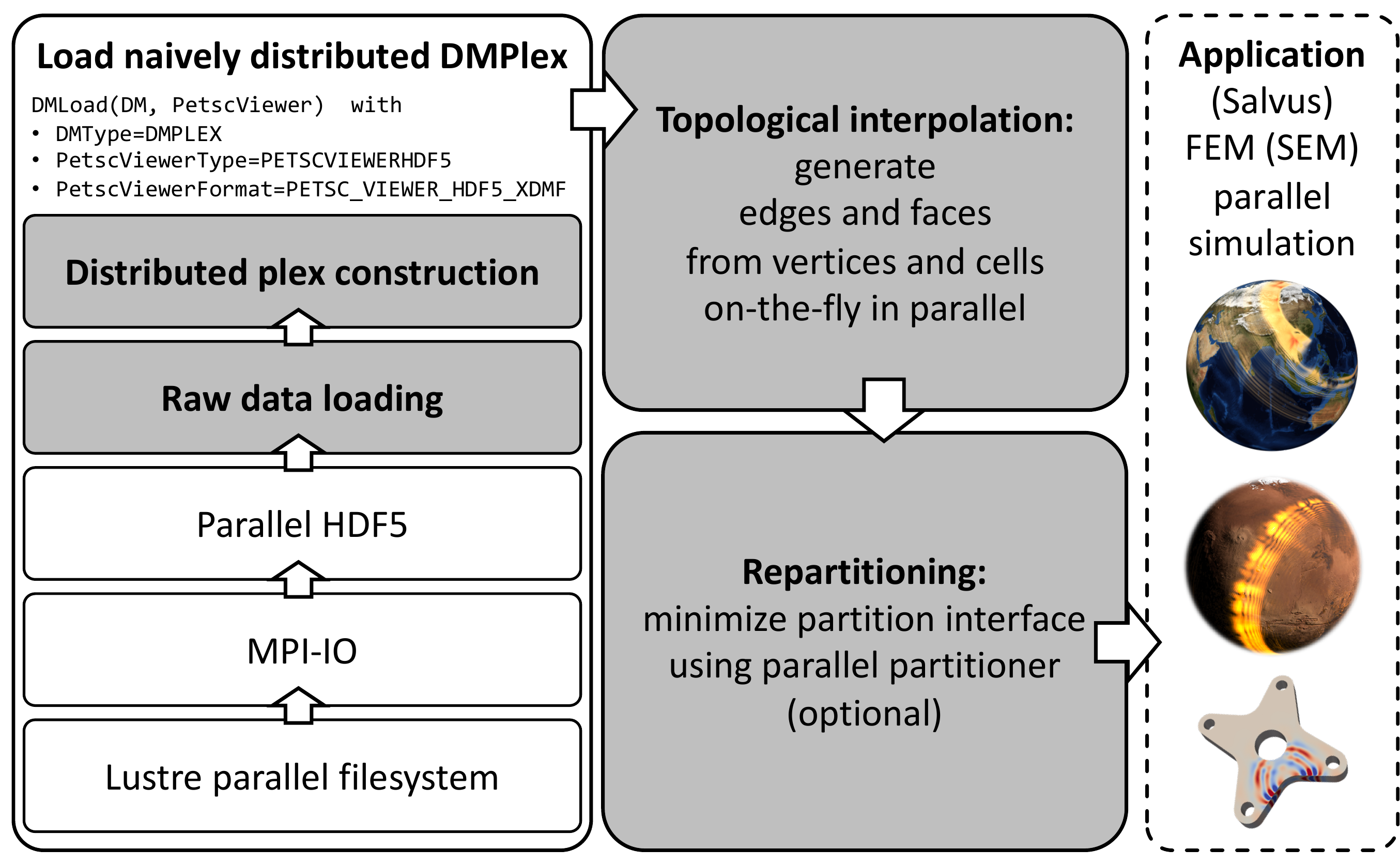}
    \caption{Schematic diagram of the parallel simulation startup. Grey stages are within PETSc scope.}
    \label{fig:startup}
\end{figure}

Before the developments of this paper, these steps were serial and only the last step, a~one-to-all distribution using a~serial partitioner such as METIS \cite{metis,karypis_fast_1998}, resulted in the distributed mesh.
This approach inevitably led to the upper limit on the mesh size due to the memory constraints of a~single node of a~cluster.
Therefore we developed a~new, completely parallel startup phase where all four steps are done in parallel right from the beginning.
This parallel startup phase is schematically depicted in \cref{fig:startup}.
We further show that even for meshes that fit into memory, parallel startup can bring significant time and energy saving.
Let us describe these stages more in detail in the following subsections.

\subsection{Raw data loading}\label{sec:raw_loading}
This stage forms the first part of our mesh reader implementation,
and consists in reading distributed raw topology and geometry data by generic index set and vector readers,
dominated by low level I/O operations.

\subsubsection{HDF5}\label{sec:hdf5}
Hierarchical Data Format 5 (HDF5)~\cite{hdf5} is a file format and library, designed to store and organize large amounts of N-dimensional array data.
It is currently supported by the HDF Group, a not-for-profit corporation whose mission is to ensure continued development of HDF5 technologies and the continued accessibility of data stored in HDF.

HDF5's file structure includes two major types of objects:
\begin{inlinenum}
\item data\-sets, multidimensional arrays of a homogeneous type;
\item groups, container structures which can hold datasets and other groups.
\end{inlinenum}
Every HDF5 file has a root group \verb|/|, under which one can add additional groups and datasets.
This results in a hierarchical, filesystem-like data format.
Resources in an HDF5 file can be accessed using the POSIX-like syntax \verb|/group1/group2/dataset|.
Metadata is stored in the form of user-defined, named attributes attached to groups and datasets~\cite{hdf5}.

HDF5 transparently handles how all the objects map to the actual bytes in the file.
HDF5 actually provides an abstracted filesystem-within-a-file that is portable to any system with the HDF5 library installed,
regardless of the underlying storage type, filesystem, or endianess.
It does automatic conversions between storage datatypes (dictated by the data file) and runtime memory datatypes
(dictated by the application)~\cite{hdf5}.

HDF5 supports parallel shared-file I/O using MPI-IO~\cite{mpi} capabilities which in turn provide scalable access to the underlying parallel filesystem such as Lustre~\cite{lustre}.
By default, HDF5 provides uniform access to all parts of the file for all processes of the communicator (passed to HDF5 using \verb`H5Pset_fapl_mpio()`).
However, since PETSc uses data parallelism, it would be very inefficient to load all data to all processes and then distribute them again.
The most important functionality in this regard are {\em hyperslabs} which can read or write to a portion of a dataset.
A hyperslab can be a logically contiguous collection of points in a dataspace, or a regular pattern of points or blocks in a dataspace.
The hyperslab can select a separate chunk of the file for each process individually by means of a rank-dependent offset.
The \verb|H5Sselect_hyperslab()| function is used for this purpose~\cite{phdf5}.

\subsubsection{XDMF}\label{sec:xdmf}
XDMF (eXtensible Data Model and Format)~\cite{xdmf} is a mesh data file format.
It distinguishes the metadata (light data) and the values themselves (heavy data).
Light data and heavy data are stored using XML and HDF5, respectively.
The data format is stored redundantly in both XML and HDF5.
There are two crucial datasets describing the mesh in a minimal sufficient way:
\begin{inlinenum}
\item \verb|<Geometry>|, a 2D dataset where each row contains coordinates of a vertex (2 or 3 scalars based on dimensionality);
\item \verb|<Topology>|, a 2D dataset where each row represents a cell, listing indices of all incident vertices.
\end{inlinenum}
Each vertex index in \verb|<Topology>| corresponds to a row index within \verb|<Geometry>|).
Both these datasets can be defined within the XML file as plain text,
or refer to a dataset path within the standalone HDF5 file (e.g. \verb|MyData.h5:/geometry/vertices|).
Both ways can be mixed within the same XDMF file,
and are both supported by widely used visualization programs such as ParaView, VisIt and EnSight.
Nevertheless, the former way is advisable only for small datasets.
We always store all data in the HDF5 file and use the XDMF file only as a descriptor.
We will refer to this as HDF5/XDMF format.

\subsubsection{PETSc data loading}\label{sec:petsc_loading}
PETSc contains a class\footnote{
  \verb|PetscViewer|
} designated for all I/O of all PETSc classes such as a vector, matrix, linear solver or DM.
The source/destination is dictated by the viewer type\footnote{
  \verb|PetscViewerType|, such as \verb|ascii|, \verb|binary|, \verb|hdf5|, \verb|socket|
} and additional properties such as the filename.
A more fine-grained control of how the object is read/viewed is accomplished with the viewer format\footnote{
  \verb|PetscViewerFormat|; e.g. \verb|ascii_info| and \verb|ascii_info_detail| print plain text information about the object with a different level of verbosity.
}.

Most relevant for this work is that \verb|DM| supports both reading and writing using a viewer\footnote{
  Using \verb|DMLoad(DM,PetscViewer)| and \verb|DMView(DM,PetscViewer)|.
}.
We have implemented a new format \verb|hdf5_xdmf| for the viewer type \verb|hdf5| to enable parallel reading and writing of the plex mesh representation\footnote{
  \verb|DM| with \verb|DMType = plex|
}
and stored in an HDF5 file with topology and geometry data compatible with XDMF\footnote{
  See \cref{sec:hdf5,sec:xdmf}.
}.
HDF5/XDMF has become the first widely used mesh format supported by PETSc which is both readable and writable in parallel.

As we are here interested in the simulation startup phase, we will now focus only on the HDF5/XDMF reader\footnote{
  \verb|DMLoad()| implementation for the combination \verb|DMType = plex|, \verb|PetscViewerType = hdf5| and \verb|PetscViewerFormat = hdf5_xdmf|.
}.
This implementation relies on lower level readers for index sets and scalar vectors\footnote{
  Classes \verb|IS| and \verb|Vec| and their methods \verb|ISLoad()| and \verb|VecLoad()|, respectively.
}.
They both read datasets from given paths within the HDF5 file\footnote{
  Using the \verb`H5Dread()` routine of HDF5, see \cref{sec:hdf5}.
} with respective HDF5 datatypes\footnote{
  \verb`H5T_NATIVE_INT` and \verb`H5T_NATIVE_DOUBLE`
}.
They make use of a hyperslab\footnote{
  See \cref{sec:hdf5}.
} reflecting the given parallel layout\footnote{
  \verb|PetscLayout|
} to divide the global dataset into local chunks along the first dimension.
The layout is either specified by user, or calculated automatically so that the chunks' lengths differ by 1 at most.
The second dimension of the dataset is interpreted as a block size,
i.e., the resulting vector is divided into equally sized shorter blocks of this size.
Blocks can have various contextual meanings such as DOFs of the same element.

The reader loads the cell-vertex topology information first\footnote{
  The \verb|<Topology>| XDMF dataset, see \cref{sec:xdmf}, loaded using \verb|ISLoad()|.
}.
Each block corresponds to an element,
and each single entry refers to one of this element's vertices using the implicit global vertex numbering.
Global indices of the blocks form an implicit global cell numbering.

Further, the geometry information is loaded into a vector\footnote{
  The \verb|<Geometry>| XDMF dataset, see \cref{sec:xdmf}, loaded using \verb|VecLoad()|.
},
whose blocks and entries represent vertices and their coordinates, respectively.
The size of all blocks is the same and corresponds to the spatial dimension of the mesh,
and global indices of the blocks form an implicit global vertex numbering.

When we read such representation in parallel, all processes load approximately equally sized, contiguously numbered, disjoint portions of vertices and cells in a single parallel I/O operation.
Note that the global vertex and cell numberings do not depend on the number of processes.

\subsection{Distributed plex construction}\label{sec:par_plex}
The raw topology and geometry data loaded in \cref{sec:raw_loading} need to be transformed into a plex representation.
This forms the second part of \verb`DMLoad()`,
and is realized by a call to a communication-bounded operation described in detail in \cref{sec:par_plex_create}.
The resulting \verb`DMPlex` instance is naively distributed with vertices and cells only.
Let us first describe a parallel star forest graph implementation,
which allows gluing together serial plexes across different processes.

\subsubsection{Star Forest}\label{sec:sf}
The Star Forest (SF)\footnote{
  In PETSc called \verb|PetscSF|~\cite{petsc-user-ref,StarForest11}.
} is a forest of star graphs in the graph theory language.
The \textit{root} point $p$ of the star is exclusively owned by a single process.
There is one and only one root per star, so we can denote the star as $p$ as well.
The \textit{leaves} of the star $p$ are shared versions of the root point $p$ on other processes.
The leaves can be represented as pairs $(r, q)$, where $r$ is a process rank, and $q$ is a remote point on the rank $r$ in this rank's own numbering.
So each star graph forms a one-to-many mapping $p \rightarrow (r_{p,0},q_{q,0}), (r_{p,1},q_{q,1}), \ldots, (r_{p,n_p},q_{q,n_p})$,
where $n_p$ is the number of the leaves for the root $p$.

Each star graph represents a shared datum such as a mesh entity or solution degree-of-freedom (DOF).
The root then represents the ``master point'' and the leaves represent ``ghost points''.
However, the SF structure represents purely a communication pattern.
Communicated data buffers can have any MPI datatype, and form parameters to SF communication routines,
rather than being statically attached to the SF root/leave points.
So a single SF instance can communicate different data of different type across its edges.
SF supports a broadcast operation from roots to leaves, as well as a reduction from leaves to roots. In addition, it
supports gather and scatter operations over the roots, inversion of the SF to get two-sided information, and a
fetch-and-op atomic update.

As we can see, the SF deals only with local numberings.
It promotes the simple design of a distributed plex,
consisting virtually of a serial plex on each rank, and a SF connecting these serial plexes together~\cite{KnepleyLangeGorman2015}.
Global numberings do not need to be maintained and are computed only if needed by the application.

\begin{figure}
    \begin{subfigure}{0.49\columnwidth}
        \centering
        \includegraphics[width=.9\linewidth]{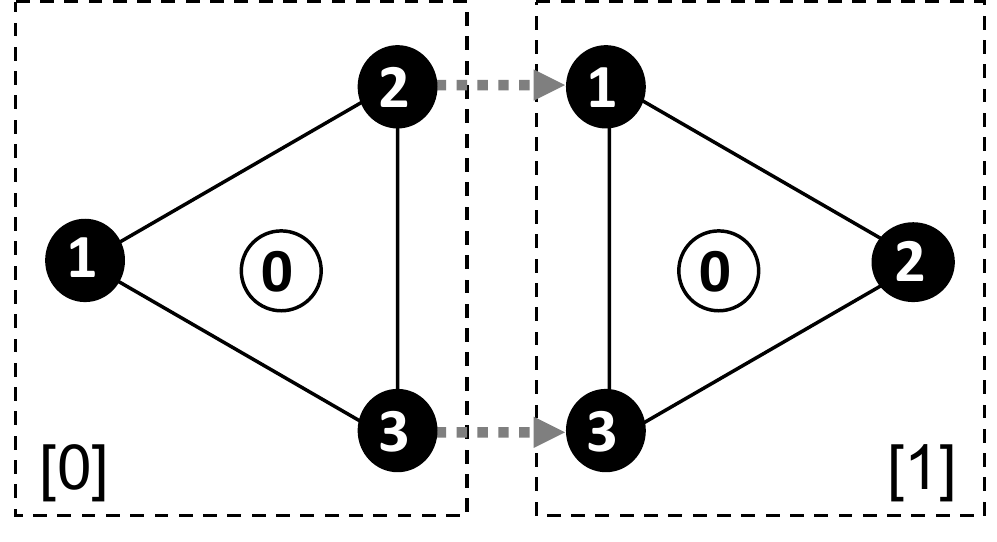}
        \caption{mesh from \cref{fig:seq_mesh_orig}\\ after distribution}
        \label{fig:par_mesh_orig}
    \end{subfigure}
    \hfill
    \begin{subfigure}{0.49\columnwidth}
        \centering
        \includegraphics[width=.9\linewidth]{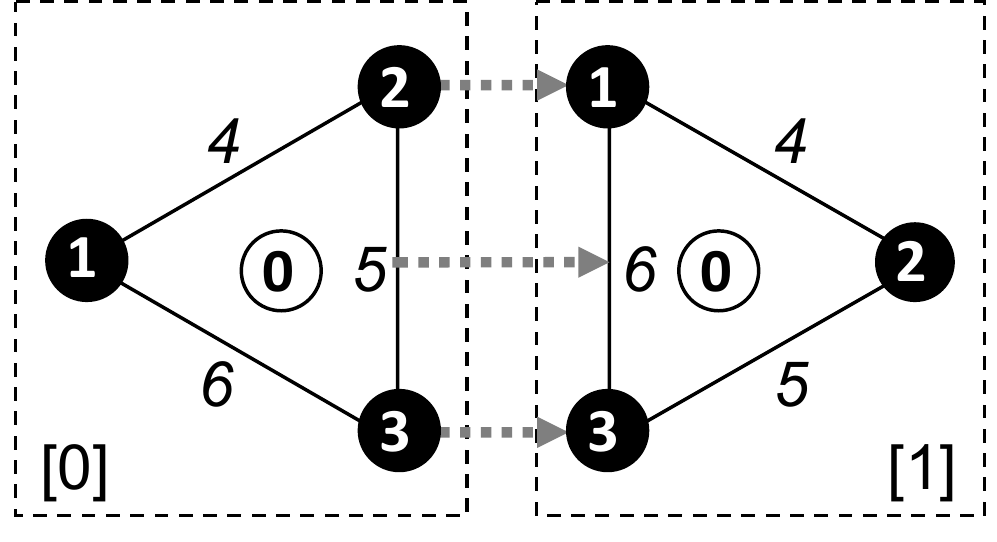}
        \caption{mesh from \subref{fig:par_mesh_orig}\\ after interpolation}
        \label{fig:par_mesh_int}
    \end{subfigure}
    \\
    \begin{subfigure}{0.49\columnwidth}
        \centering
        \includegraphics[width=.9\linewidth]{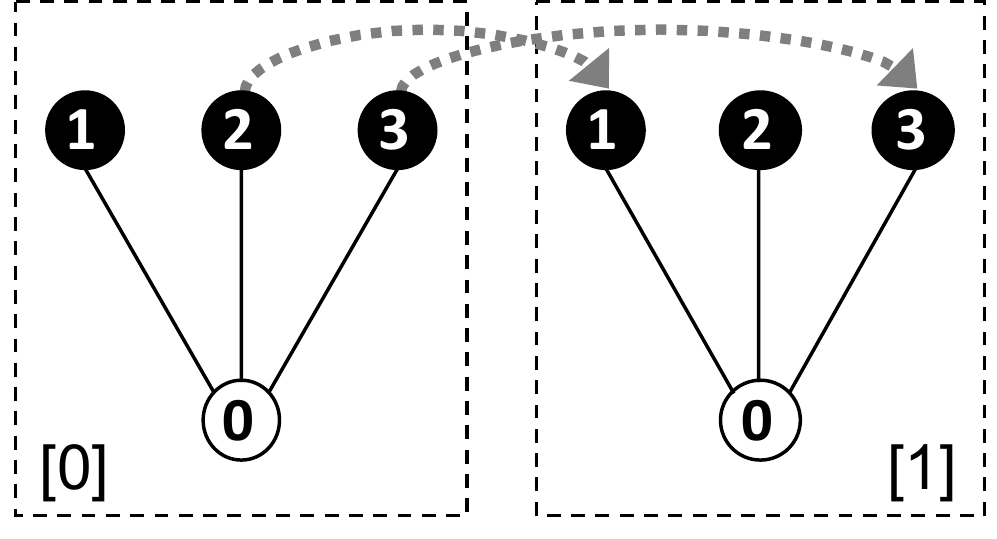}
        \caption{plex representation of \subref{fig:par_mesh_orig}}
        \label{fig:par_mesh_plex}
    \end{subfigure}
    \hfill
    \begin{subfigure}{0.49\columnwidth}
        \centering
        \includegraphics[width=.9\linewidth]{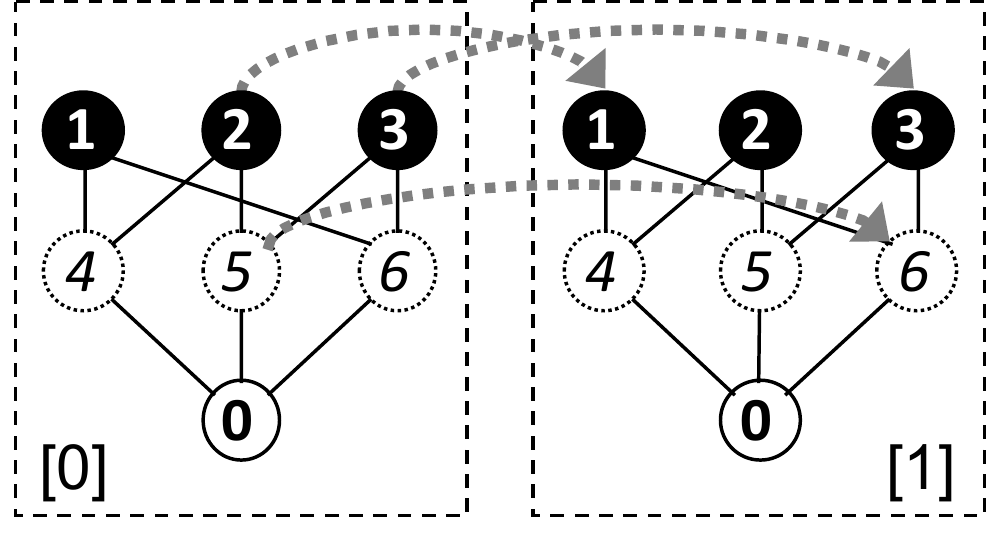}
        \caption{plex representation of \subref{fig:par_mesh_int}}
        \label{fig:par_mesh_int_plex}
    \end{subfigure}
    \caption{Parallel DMPlex. Grey dotted arrows denote \texttt{sfPoint}.}
    \label{fig:par_mesh}
\end{figure}

\subsubsection{Construction of distributed plex from raw data}\label{sec:par_plex_create}
Once the raw topology and geometry data is loaded as described in \cref{sec:petsc_loading},
vertex coordinates and the cells are initially distributed over ranks independently.
Concerning the cell-vertex topology information, rank $p$ gets $N^{(p)}_C \times c$ global vertex numbers (integers),
where $N^{(p)}_C$ is the number of cells on rank $p$, and $c$ is the number of vertices per cell\footnote{
  We assume here meshes with a uniform cell type.
}.

Concerning the geometry information, rank $p$ gets $N^{(p)}_V \times d$ coordinates (real numbers),
where $N^{(p)}_V$ is the number of vertices on rank $p$, and $d$ is the mesh dimension\footnote{
  For example, if we have 20 vertices
  in 3D and 4 ranks, each rank might get 15 doubles, namely the coordinates for 5 contiguously numbered vertices,
  regardless of the cells it has been assigned to.
}.
The layout of the geometry information dictates the initial vertex distribution.

Our goal now is to create a distributed plex from these inputs.
The original cell partitions are kept, and a serial plex object is built from each of them.
These serial plexes are then combined together by the parallel SF object \verb|sfPoint| as shown in \cref{fig:par_mesh}.
Let us now describe this construction process\footnote{
  Implemented in the PETSc function \verb|DMPlexCreateFromCellListParallel()|.
} in detail.

In order to get partition-wise complete topological information,
the vertices need to be redistributed so that each partition possesses all incident vertices of its cells locally.
The redistribution involves duplication of vertices on partition boundaries.
So we work here with two sets of vertices: the original {\em global vertices} using the global numbering and initial distribution,
and the new {\em local vertices} with the local numbering and localized distribution.

We iterate through the cell-vertex topology information and put the global vertices as keys into a hash set\footnote{
  By a hash set, we mean a hash table without values, only keys.
}.
One this loop is finished, the size of the hash set equals the number $n$ of local vertices a rank will own.
Entries of the hash set are extracted into an array $v$ containing $n$ locally unique global vertices.
We sort $v$, and the positions $0, \ldots, n-1$ can then be taken as the local vertex numbering.
We can use it along with the initial distribution\footnote{
  Represented by a \verb|PetscLayout| object in PETSc.
} to construct a~SF instance \verb|sfVert|\footnote{
  Using the PETSc function \verb|PetscSFSetGraphLayout()|.
}.
This SF realizes the mapping between the two sets of vertices.
We can use it to broadcast the coordinates\footnote{
  \verb|PetscSFBcastBegin()|/\verb|PetscSFBcastEnd()|
},
so that we get the initial geometrical information correctly mapped to the local vertices.

Further, we construct cones of the DAG points representing cells.
We simply translate the global vertices of each cell vertex tuple in the cell-vertex topology information to the local vertices by looking up position in the array $v$\footnote{
  The array is sorted so we can use bisection searching.
  Since \verb|DMPlex| uses unique DAG point numbering, and DAG point numbers $0, \ldots, (N^{(p)}_C - 1)$ are assigned to cells,
  the local vertices are additionally shifted by the constant $N^{(p)}_C$.
}.
Once we have the complete cone information, we symmetrize it to get the support information\footnote{
  \verb|DMTPlexSymmetrize()|},
and construct strata\footnote{
  \verb|DMPlexStratify()|}.
Both these operations have linear complexity.

We can use \verb|sfVert| once again to construct \verb|sfPoint|, the SF which ``holds together'' the distributed plex
by describing sharing of the local vertices between processes\footnote{
  It connects adjacent local vertices on different ranks, which correspond to the same vertex before distribution,
  i.e. have the same coordinates, as shown in \cref{fig:par_mesh}.
}.
In order to construct \verb|sfPoint|, we mainly need to determine a unique owner for each vertex.
We first construct an array which holds remote vertices $(r, q)$ for each local vertex\footnote{
  As in \cref{sec:sf}, $q$ is the remote vertex on the rank $r$ in this rank's own numbering.
}.
Then we reduce this array from the leaves to the roots of \verb|sfVert|, yielding $(\hat{r}, \hat{q})$ at each root, where $\hat{r}$ is the maximum $r$ among leaves of this root\footnote{
  \verb|PetscSFReduceBegin|/\verb|PetscSFReduceEnd| with the \verb|MPI_MAXLOC| reduce operation can be employed.
  We could also use minimum, or some other non-ambiguous choice.
}.
This pair is then broadcast back to the leaves\footnote{
  \verb|PetscSFBcastBegin()|/\verb|PetscSFBcastEnd()|
}, giving the needed unique owner for each shared vertex.
The owner $(\hat{r}, \hat{q})$ is then easily translated to a root of \verb|sfPoint|, and the rest of adjacent vertices to leaves.

\subsection{Parallel topological interpolation}\label{sec:par_interpolation}
Since the steps above lead to a distributed plex, we need a parallel version of the topological interpolation (\cref{sec:interpolation}).
It consists of the serial interpolation (in-memory computations) and a small communication.

The first step consists in applying the sequential topological interpolation (\cref{sec:interpolation}) and cone orientation (\cref{sec:orient}) on each rank independently.
Then we must alter \verb|pointSF|, the SF instance
which identifies points owned by different processes, or \textit{leaf} points\footnote{
  Let us remind our {\em points} are {\em DAG points} which represent mesh entities of all codimensions.
  In this particular case, we speak about boundary entities whose codimension is up to 1.
  For example, vertices, edges and faces in the 3D case.
}.
We mark all leaf points which are adjacent to another ghost point as candidates. These candidate
points are then gathered to \textit{root} point owners\footnote{
  \verb|PetscSFGatherBegin()|/\verb|PetscSFGatherEnd()|
}.
For each candidate, the root
checks for each point in the cone that either it owns that point in the SF or it is a local point. If so, it claims
ownership. These claims are again broadcast, allowing a new SF to be created incorporating the new edges and faces.

\begin{figure}
    \begin{subfigure}{0.49\columnwidth}
        \centering
        \includegraphics[width=.9\linewidth]{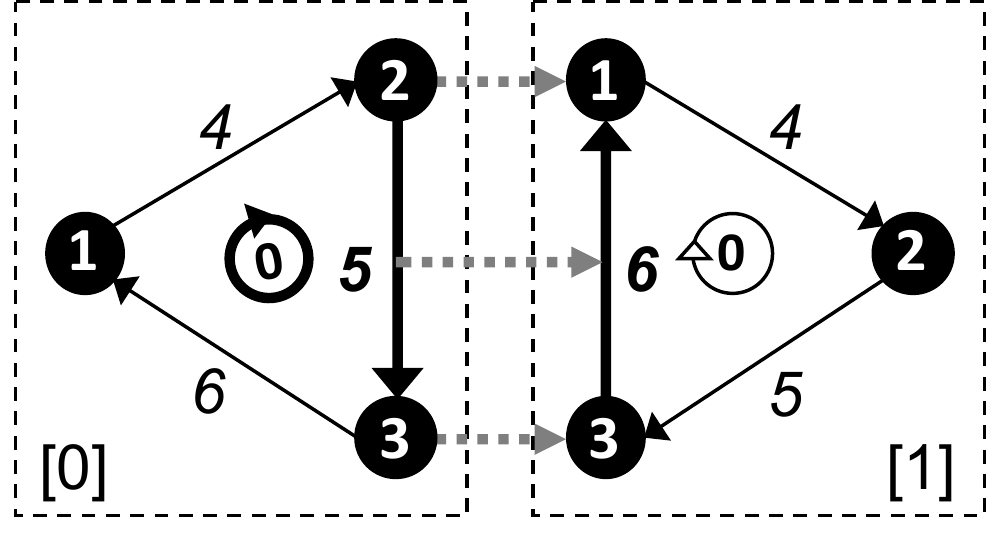}
        \caption{mesh from \cref{fig:par_mesh_int}\\with non-conforming cone orientation}
        \label{fig:par_ornt_wrong}
    \end{subfigure}
    \hfill
    \begin{subfigure}{0.49\columnwidth}
        \centering
        \includegraphics[width=.9\linewidth]{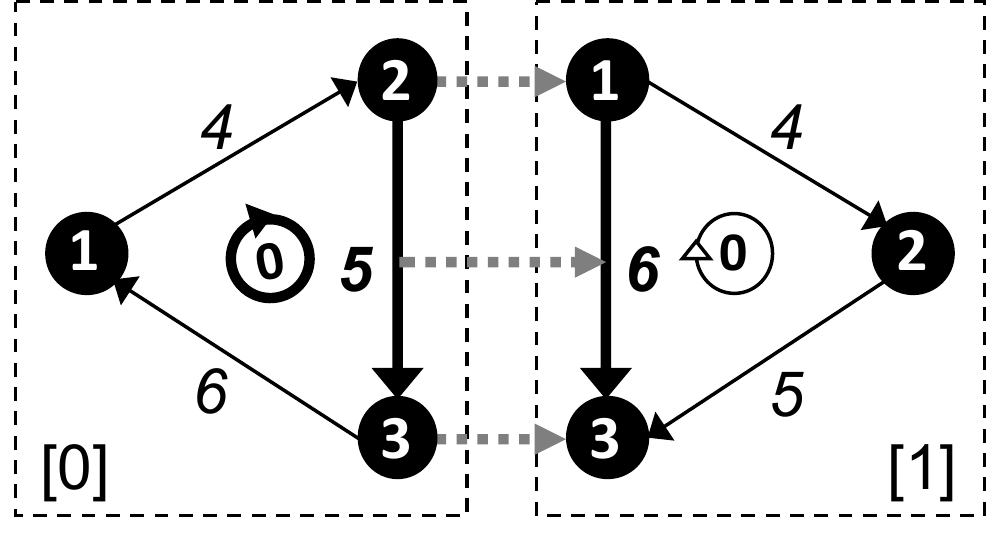}
        \caption{mesh from \cref{fig:par_mesh_int}\\with conforming cone orientation}
        \label{fig:par_ornt_correct}
    \end{subfigure}
    \\
    \begin{subfigure}{0.49\columnwidth}
        \centering
        \includegraphics[width=.9\linewidth]{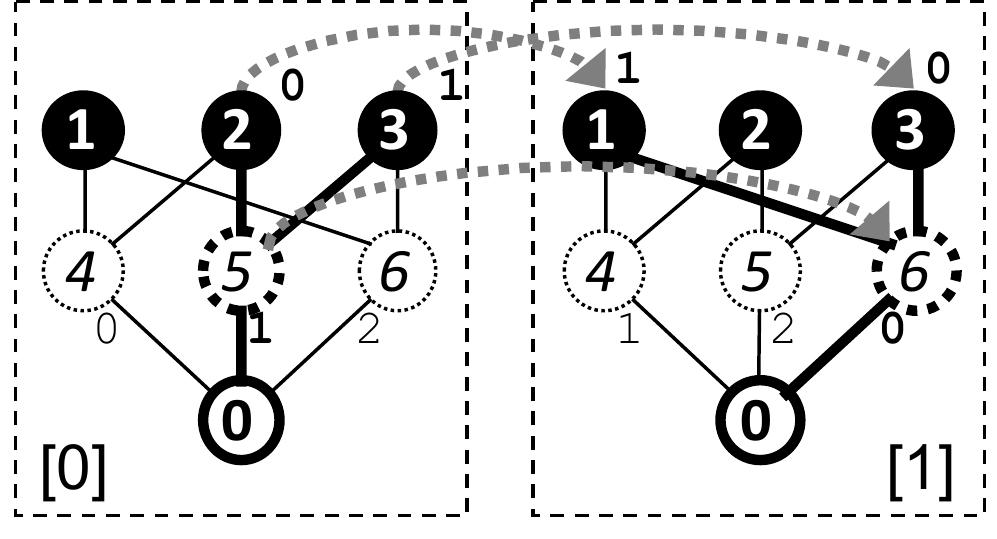}
        \caption{plex representation of \subref{fig:par_ornt_wrong};\\$O([1]0,0)=0$}
        \label{fig:par_ornt_wrong_plex}
    \end{subfigure}
    \hfill
    \begin{subfigure}{0.49\columnwidth}
        \centering
        \includegraphics[width=.9\linewidth]{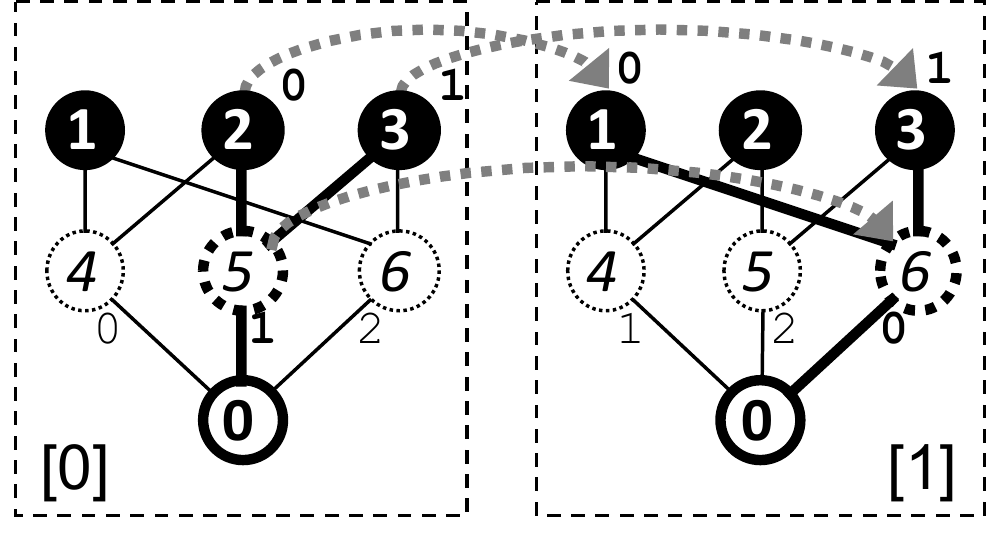}
        \caption{plex representation of \subref{fig:par_ornt_correct};\\$O([1]0,0)=-2$}
        \label{fig:par_ornt_correct_plex}
    \end{subfigure}
    \caption{Parallel DMPlex with cone points order and orientation.}
    \label{fig:par_ornt}
\end{figure}

The cone orientation has been done on each rank independently, and hence it is only partition-wise correct.
However, we have not yet handled the following assumption:
If interface edges/faces owned by different ranks represent the same geometrical entity,
i.e., they are connected by \verb|pointSF|
like edges $[0]5$ and $[1]6$ in \cref{fig:par_ornt},
they must have a conforming order of cone points ($[r]$ means ownership by rank $r$).
This requirement can be written more rigorously as an implication
\begin{align}
\begin{rcases}
p_0 &\to p_1 \\
C(p_0) &= (q_{0,0},\ldots,q_{0,n-1})\\
C(p_1) &= (q_{1,0},\ldots,q_{1,n-1})
\end{rcases}
\Rightarrow
\begin{cases}
q_{0,0} &\to q_{1,0}\\
&\cdots\\
q_{0,n-1} &\to q_{1,n-1},
\end{cases}
\end{align}
using notation from \cref{sec:orient}
and relation $\to$ meaning a connection via \verb|pointSF|.

In \cref{fig:par_ornt_wrong,fig:par_ornt_wrong_plex} this assumption is violated for the edges $[0]5$ and $[1]6$.
They are flipped to each other, more rigorously speaking \verb|pointSF| connects the edge and its incident vertices
\begin{align*}
  [0]5 \to [1]6,\qquad  [0]2 &\to [1]1,\qquad  [0]3 \to [1]3,
\end{align*}
but the order of cone points does not conform,
\begin{align*}
  [0]2 = C([0]5, 0) &\not\to C([1]6, 0) = [1]3,\\
  [0]3 = C([0]5, 1) &\not\to C([1]6, 1) = [1]1.
\end{align*}
This would lead to incorrect PDE solution if the used discretization method makes use of the edge.

In order to satisfy this requirement, and additional synchronization of the interface cones must be carried out.
We start by synchronization of the interface cone point numbering.
Let us remind that \verb|pointSF| is a one-sided structure, so only the origins of the arrows can be found directly.

Let us assume rank $r_0$, its edge/face $[r_0]p$,
and that there is a \verb|pointSF| arrow pointing {\em from} $[r_0]p$ to some $[r_1]p$
If we detect an arrow directed {\em from} the cone point $C([r_0]p,c) \to [r_1]q_c$,
we set $\ROOT([r_0]p,c)=(r_1,q_c)$, otherwise $\ROOT([r_0]p,c)=(r_0,C([r_0]p,c))$.
This $\ROOT([r_0]p,c)$ is sent to $r_1$ using \verb|PetscSFBcastBegin/End()|,
and stored at the destination rank as $\LEAF([r_1]p,c)$.
This is done for each rank, each point in $\str(h)$, $h>0$, and each $c=0,1$.

Now from the rank $r_1$ view, it has for $c=0,1$ point $[r_1]p$,
$\ROOT([r_1]p,c)$ and the received $\LEAF([r_1]p,c)$.
If
$\ROOT([r_1]p,c) = \LEAF([r_1]p,c)$
does not hold for both $c=0,1$,
we must rotate and/or flip the cone so that this condition gets  satisfied.
In that case we must also update $O(s)$ for all $s \in S([r_1]p)$
accordingly to compensate the change of cone order.

We can see that the orientation synchronization heavily relies on the correct \verb|pointSF|.
This is why it must be processed first.

\subsection{Redistribution}
We now have a correct distributed \verb`DMPlex` instance representing all codimensions.
However, the cell distribution is naive; it is load balanced with respect to the size of partitions but the partition shape is not optimized.
It can optionally be further improved using a parallel partitioner to minimize the partition interfaces and hence reduce the halo communication in the subsequent application computation.
PETSc offers interfaces to ParMETIS~\cite{parmetis,karypis_fast_1998,karypis_parallel_1998} and PT-Scotch~\cite{ptscotch}.
For this paper, we have always used ParMETIS.
Note that thanks to the nature of the plex representation, no explicit dual graph computation is needed in order to partition cells \cite{knepley_mesh_2009}.
Further details of the redistribution stage are described, e.g., in \cite{KnepleyLangeGorman2015}.

\section{Seismic wave propagation modeling}\label{sec:salvus}
As a representative use case and benchmarking tool for the new parallel simulation startup phase described above in \cref{sec:startup},
we use an implementation of the spectral-element method (SEM).

Although originally developed for applications in fluid dynamics~\cite{Patera84}, 
continuous-Galerkin SEM on hexahedral elements has emerged as the de-facto standard for global-scale seismic wave simulations~\cite{Peter_2011,Ferroni_2017,Fichtner_book}.
SEM is a~high-order finite-element method with very low dispersion and dissipation errors~\cite{Ferroni_2017}.
The choice of the Gauss-Lobatto-Legendre collocation points for the interpolating Lagrange
polynomials naturally yields a diagonal mass matrix, which enables the use of explicit
time stepping schemes.
A~second-order Newmark time-stepping additionally allows us to compute coupling terms
along any solid-fluid interfaces without the need to solve a linear system~\cite{NissenMeyer_2008}.
Furthermore, the tensorized structure of the finite-element basis on
hexahedral elements allows for efficient computations of internal forces.
These element-wise operations can be formulated as dense matrix–matrix products,
making the method suitable for the current generation of SIMD computing architectures.

Salvus \cite{Afanasiev_2019} contains a flexible implementation of SEM,
separating the wave propagation physics, the spatial discretization and finite-element shape mappings into distinct and functionally orthogonal components.
It uses modern C++ features to ensure that runtime performance is not affected.
It is parallelized using MPI and GPU-accelerated with CUDA.
PETSc DMPlex (\cref{sec:dmplex}) is used for mesh management so the developments of this paper can be directly applied.

\section{Performance results}
This section presents scalability tests of the new parallel simulation startup phase (\cref{sec:startup})
used within seismic wave propagation modeling (\cref{sec:salvus}).

\subsection{Hardware}
All benchmarks were run at Piz Daint, the flagship system of the Swiss National Supercomputing Centre (CSCS).
Piz Daint consists of 5704 12-core Cray XC50 nodes with GPU accelerators, and 1813 36-core Cray XC40 nodes without accelerators.
All benchmarks presented in this paper ran on the XC50 nodes.
Each of them is equipped with one 12-core Intel Xeon E5-2690 v3 (Haswell) processor, one NVIDIA Tesla P100 16GB GPGPU accelerator and 64 GB RAM.
Piz Daint has 8.8 PB shared scratch storage with the peak performance of 112 GiB/s.
It is implemented using Cray Sonexion 3000~\cite{sonexion3000} scale-out storage system equipped with the Lustre parallel file system~\cite{lustre}, 40 object storage targets (OSTs) and 2 metadata servers (MDSs).

\subsection{Middleware (Lustre, MPI-IO, HDF5) settings}
We always used the single shared file approach, i.e., every process reads its disjoint chunk from the common file.
There are many good reasons for such choice, such as reduction of metadata accesses,
but the main reason is the flexibility in number of processes using the same file,
unlike the file-per-process approach.
As for Lustre file system settings, we used stripe count of 40 (maximum on Piz Daint) and stripe size of 16 MB.
Regarding HDF5/MPI-IO, we always non-collective reading.
We tested also collective reading with various numbers of aggregators (MPI-IO hint \verb|cb_nodes|) but never saw any significant benefit.
The raw file reading always took less than 2 seconds; we cannot exclude that some scenarios with much bigger files and/or node counts could require more deliberate settings but such scenarios are irrelevant within the context of this paper.

\subsection{Cube benchmark}
\begin{figure}
    \begin{subfigure}{0.49\columnwidth}
        \centering
        \includegraphics[width=\linewidth]{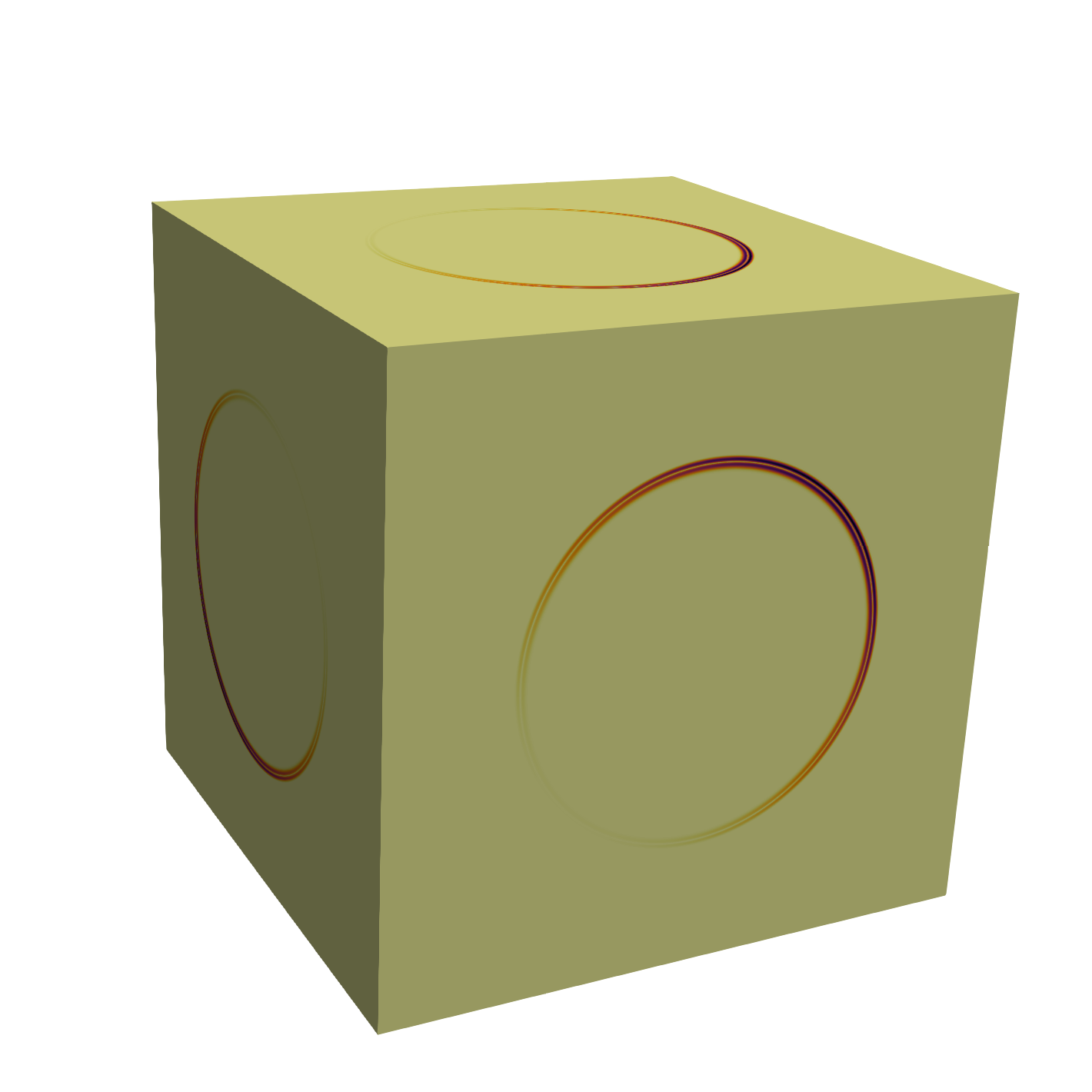}
        \caption{$t$ = 0.1 s}
        \label{fig:cube01}
    \end{subfigure}
    \hfill
    \begin{subfigure}{0.49\columnwidth}
        \centering
        \includegraphics[width=\linewidth]{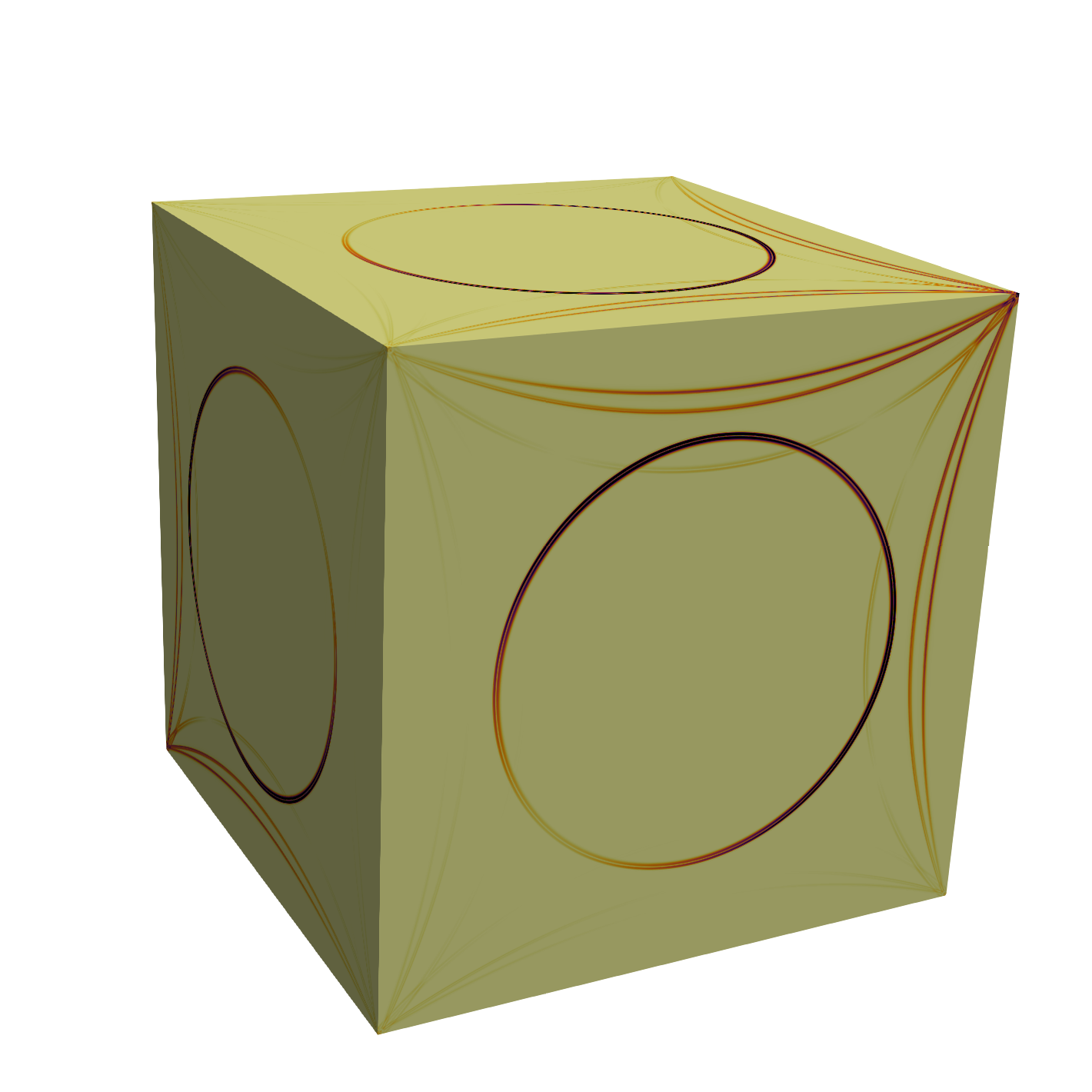}
        \caption{$t$ = 0.15 s}
        \label{fig:cube02}
    \end{subfigure}
    \caption{Cube benchmark with a moment-tensor type source located at the center of the domain.
      Isotropic elastic material model without attenuation was applied.
      The mesh size was 256 million elements, and 512 Piz Daint nodes (6144 CPUs, 512 GPUs) were used for the computation.
      The normalized magnitude of the velocity vector at time $t$ is visualized.
      In \subref{fig:cube01}, we can see the P-wave propagating from the source, while \subref{fig:cube02} shows the S-wave and reflections of the P-wave from the cube boundary.
    }
    \label{fig:cube}
\end{figure}

Our performance benchmark consists in elastic wave propagation in homogeneous isotropic media from a point source in a cubic geometry.
The cube is discretized into equally sized hexahedral cells, handled as an unstructured mesh.
Each cell hosts a 4th-order spectral element with 125 spatial DOFs.
Since a 3D vector equation is solved, this results in 1125 field variables per element, together representing acceleration, velocity, and displacement.
\Cref{fig:cube} illustrates the solution of the benchmark problem at two different timesteps.
\cref{tab:file_sizes} summarizes dimensions of the stored topology and geometry datasets and resulting file sizes for different numbers of elements in x-direction (NEX).
Our sequence of NEX was chosen so that the total number of elements (NE) of each successive mesh is approximately doubled, starting at 8 million.

\begin{figure}
    \centering
    \begin{subfigure}{.49\columnwidth}
        \includegraphics[width=\linewidth]{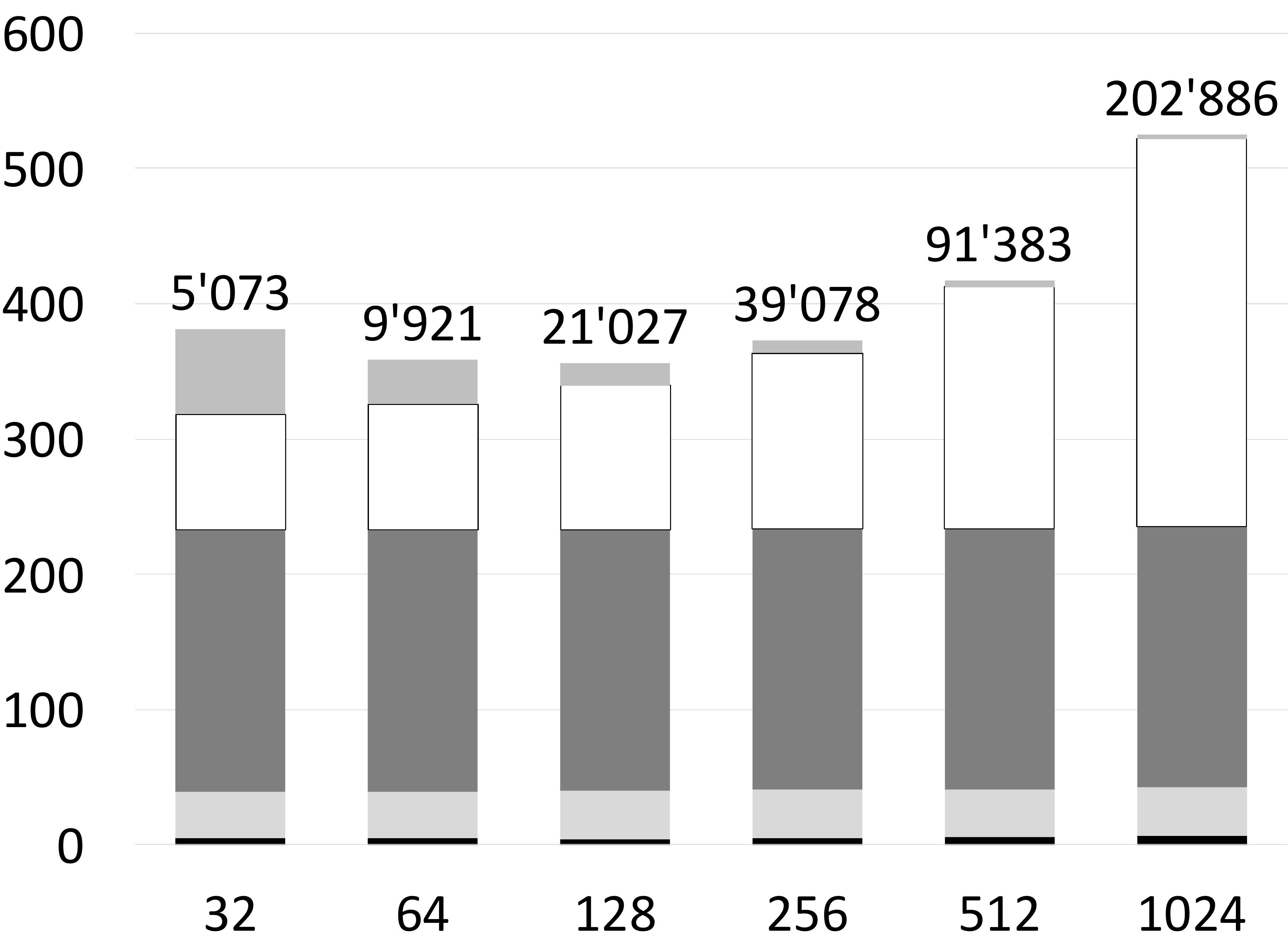}
        \caption{serial startup}
        \label{fig:scal_016m_serial}
    \end{subfigure}
    \hfill
    \begin{subfigure}{.49\columnwidth}
        \includegraphics[width=\linewidth]{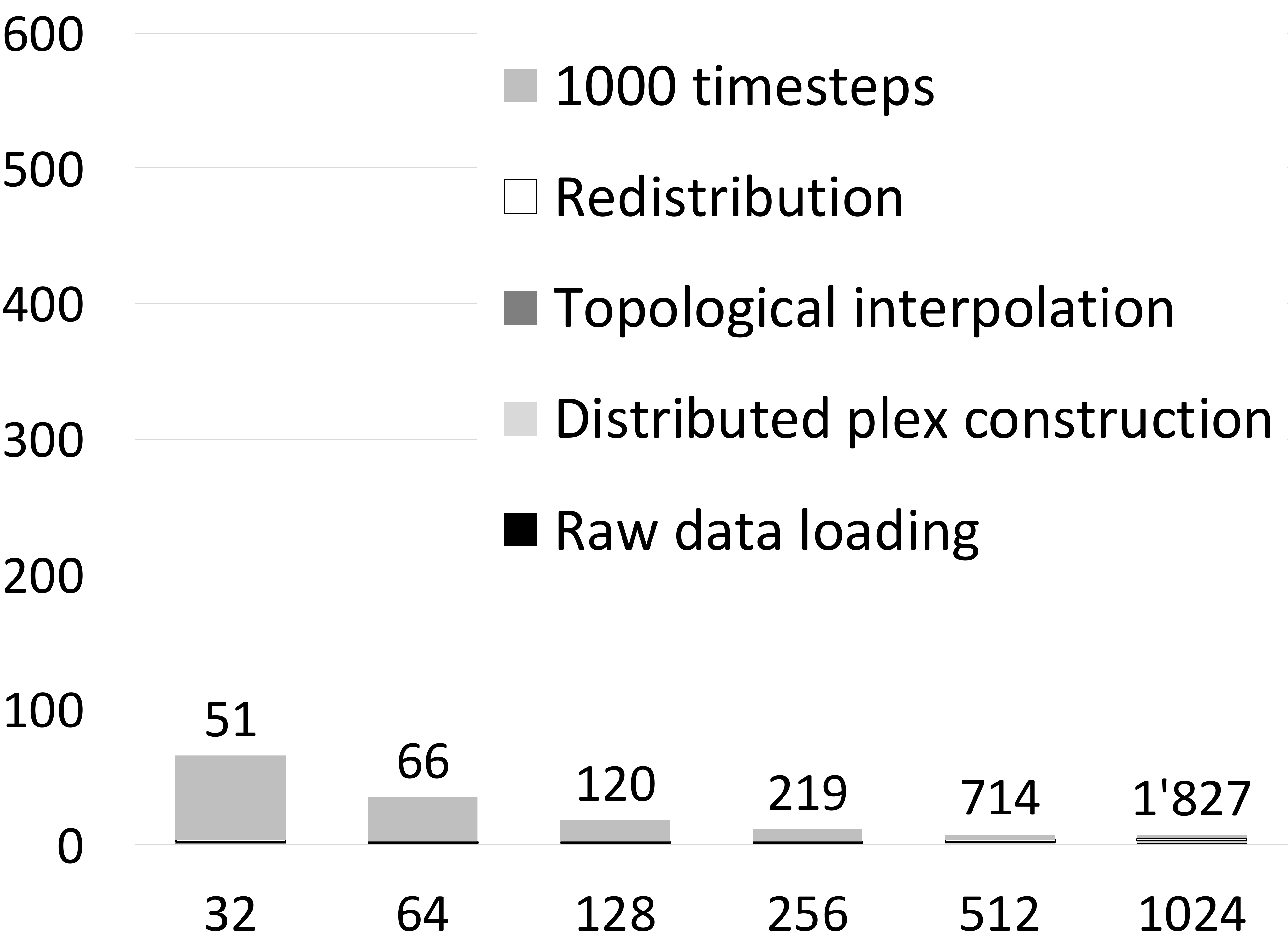}
        \caption{parallel startup}
        \label{fig:scal_016m_600s}
    \end{subfigure}
    \caption{
      {\bf Cube benchmark.
      Strong scalability, serial/parallel startup, 16 million mesh elements.}
      The serial and parallel simulation startup phase are compared with each other and 1000 steps of the Salvus time loop in terms of wall time.
      Approximate wall time for any different number of timesteps can be obtained using simple proportionality.
      The number of timesteps in production simulations varies, but generally 30'000 or more are required.
      The particular startup stages are described in \cref{sec:startup}.
      {\bf X-axis}: number of Piz Daint nodes, each equipped with 12 cores and 1 GPU per node.
      {\bf Y-axis}: wall time in seconds.
      {\bf Labels above bars}: the number of timesteps that take the same wall time as the startup phase.
      {\bf Order of colors} in the bars is the same as in the legend.
    }
    \label{fig:scal_016m_serial_vs_parallel}
\end{figure}

We present performance of the new parallel startup phase in several graphs.
Graphs in \cref{fig:scal_016m_serial_vs_parallel} present strong scalability for the mesh size of 16 million elements,
and serve mainly for comparison of the new parallel startup with the original serial startup.
16 million is an upper bound for the mesh size for the serial startup imposed by the memory of a single Piz Daint Cray XC50 node.
Overcoming this limit is for us the most important achievement of the startup phase parallelization. 
However, obviously the performance improvement is very significant as well.
At 1024 nodes, the serial startup takes an amount of wall time equivalent to 202'886 timesteps in the subsequent time loop,
whereas the parallel startup takes only 1827 timesteps,
which means 111x speedup.
The number of timesteps in production simulations varies, but generally 30'000 or more are required.

\begin{figure}
    \begin{subfigure}{.45\columnwidth}
        \includegraphics[width=\linewidth]{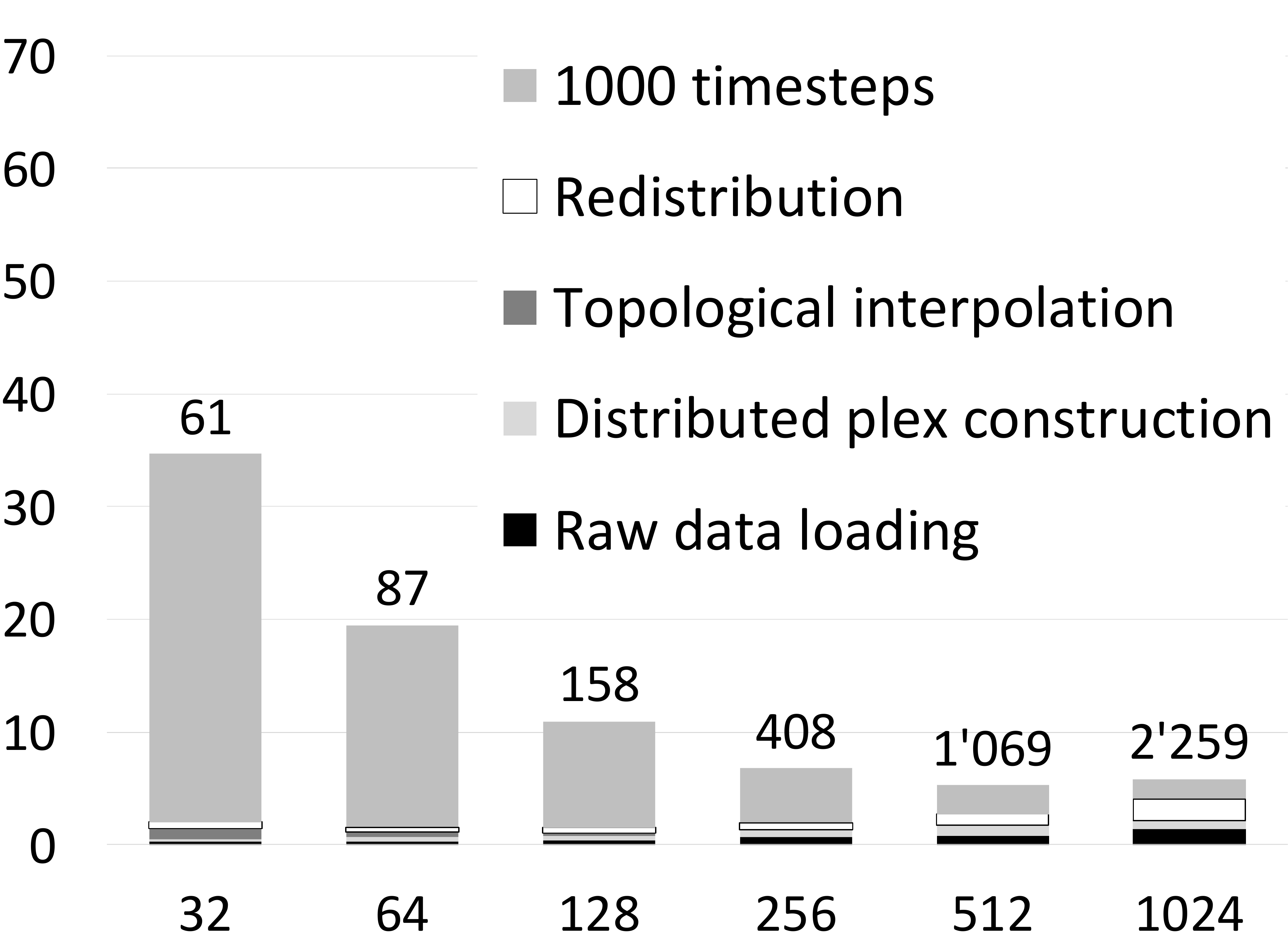}
        \caption{8 million elements}
        \label{fig:scal_008m}
    \end{subfigure}
    \hfill
    \begin{subfigure}{.45\columnwidth}
        \includegraphics[width=\linewidth]{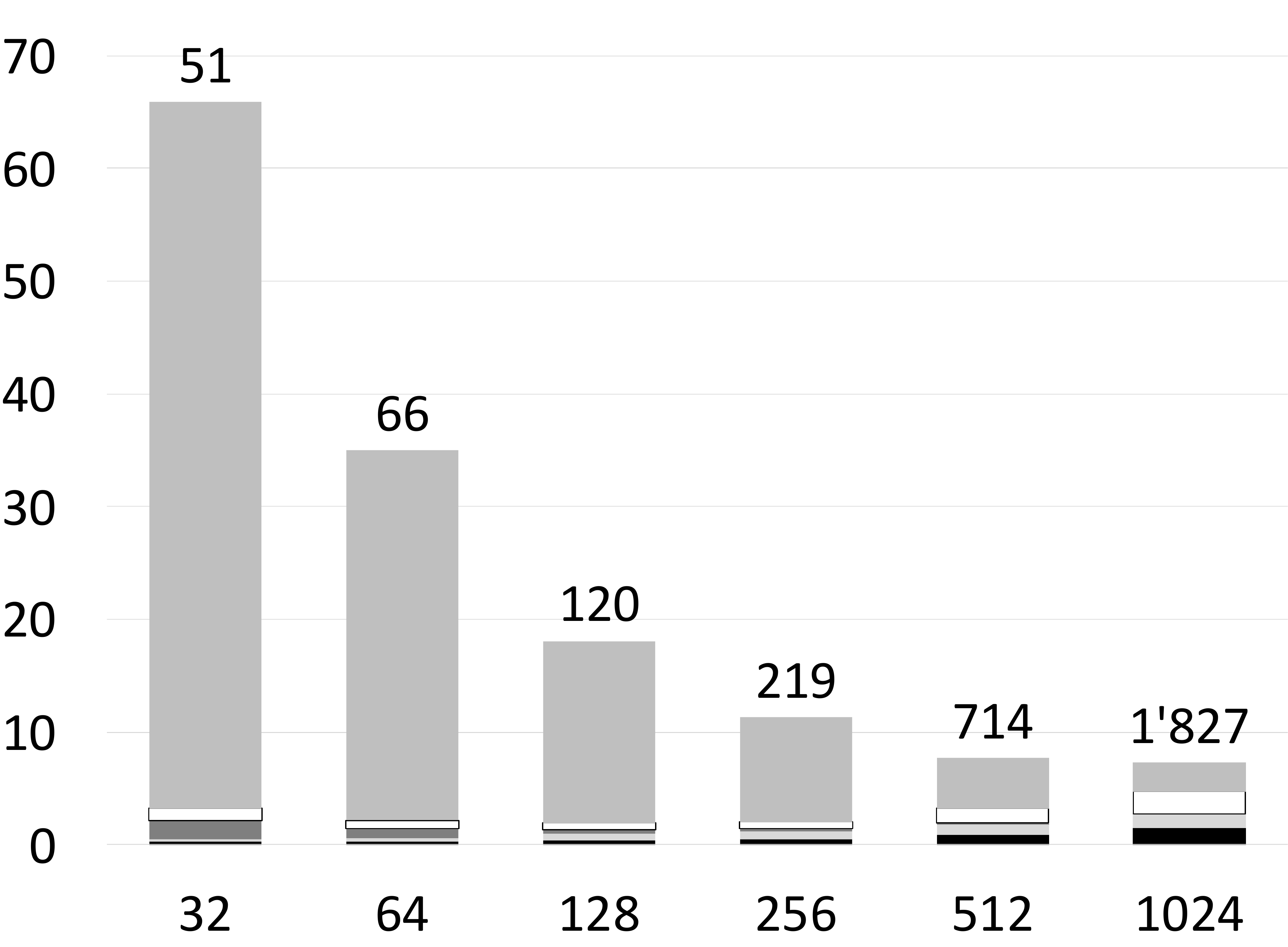}
        \caption{16 million elements}
        \label{fig:scal_016m}
    \end{subfigure}

    \begin{subfigure}{.45\columnwidth}
        \includegraphics[width=\linewidth]{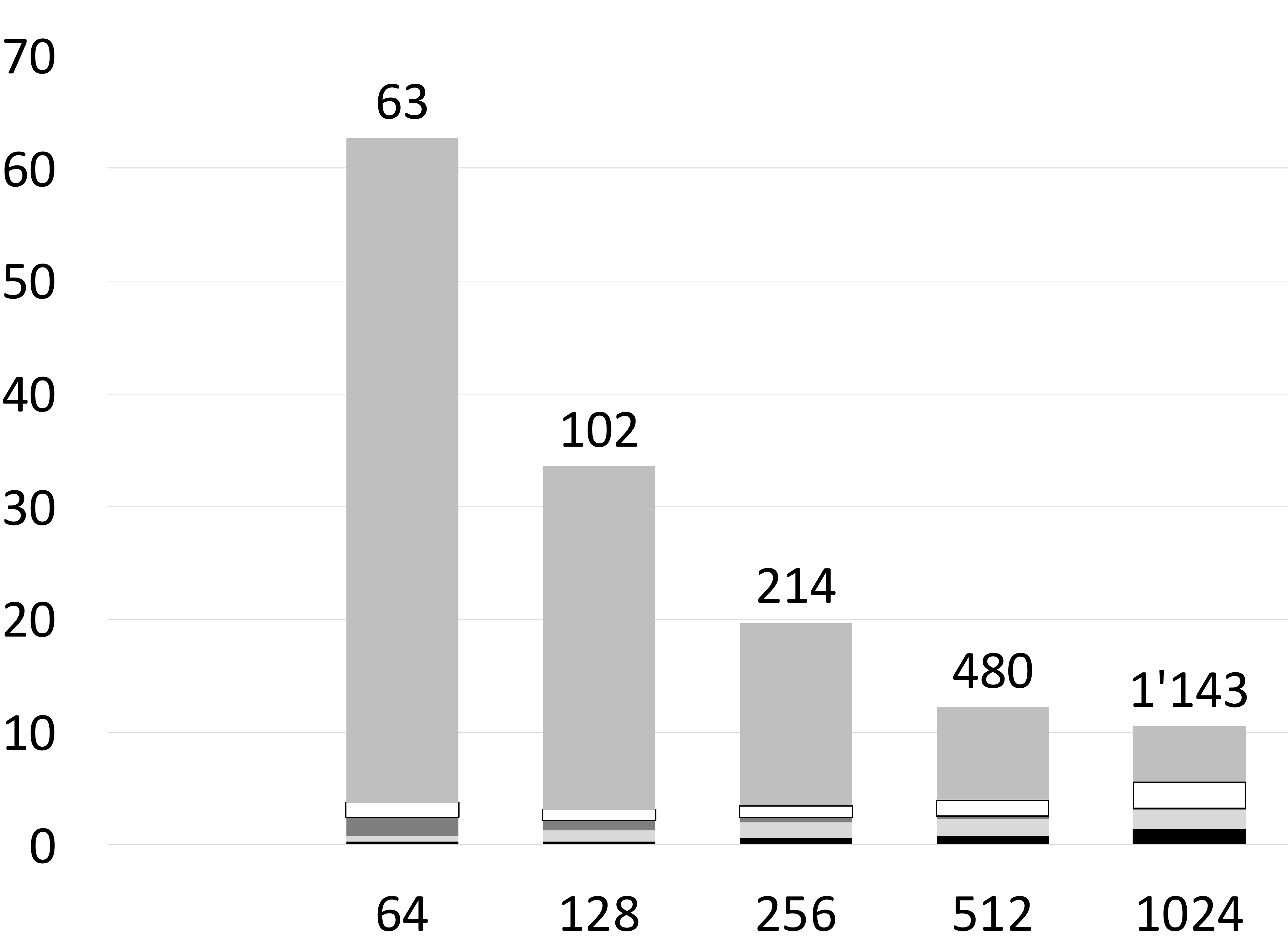}
        \caption{32 million elements}
        \label{fig:scal_032m}
    \end{subfigure}
    \hfill
    \begin{subfigure}{.45\columnwidth}
        \includegraphics[width=\linewidth]{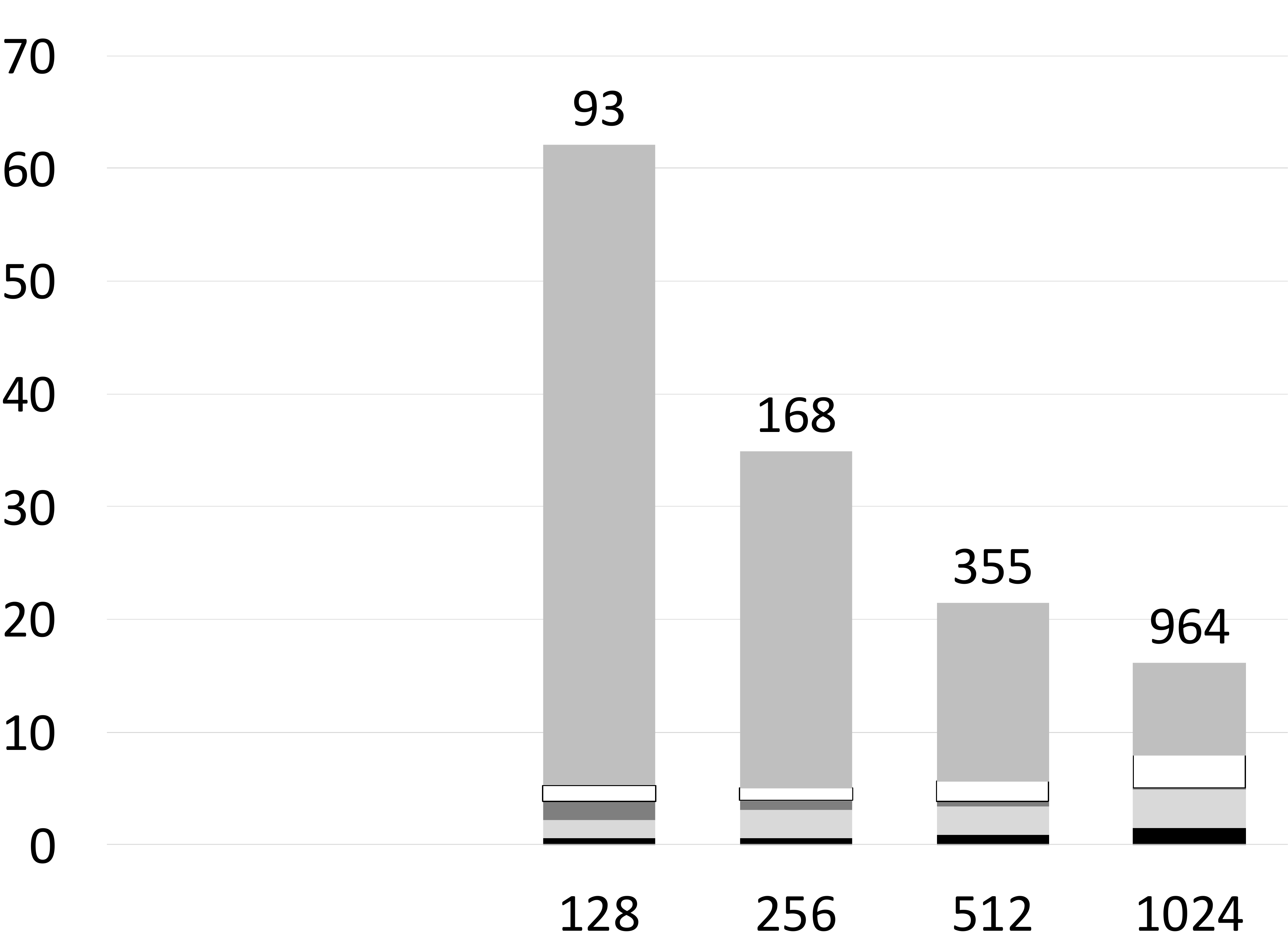}
        \caption{64 million elements}
        \label{fig:scal_064m}
    \end{subfigure}

    \begin{subfigure}{.45\columnwidth}
        \includegraphics[width=\linewidth]{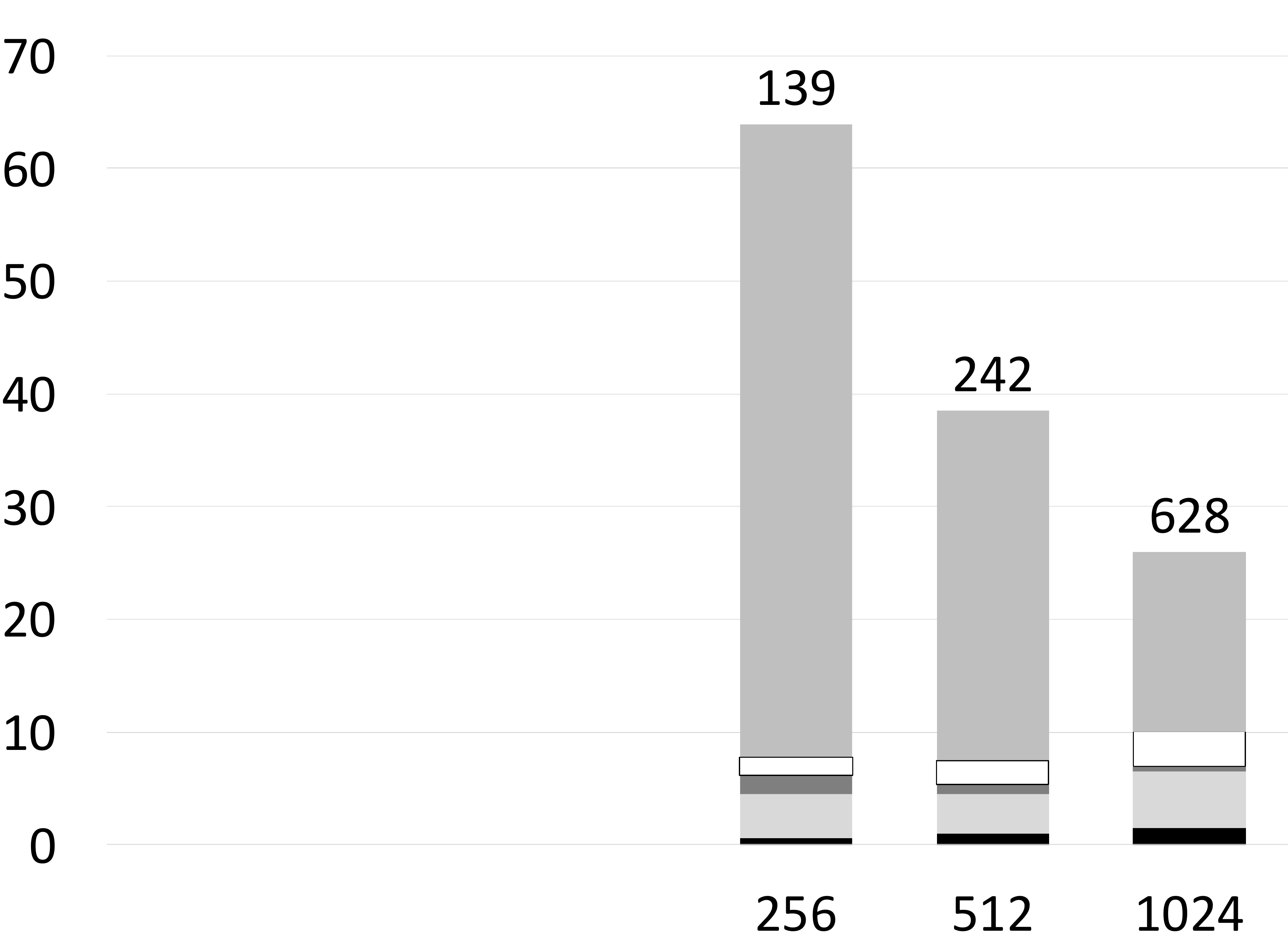}
        \caption{128 million elements}
        \label{fig:scal_128m}
    \end{subfigure}
    \hfill
    \begin{subfigure}{.45\columnwidth}
        \includegraphics[width=\linewidth]{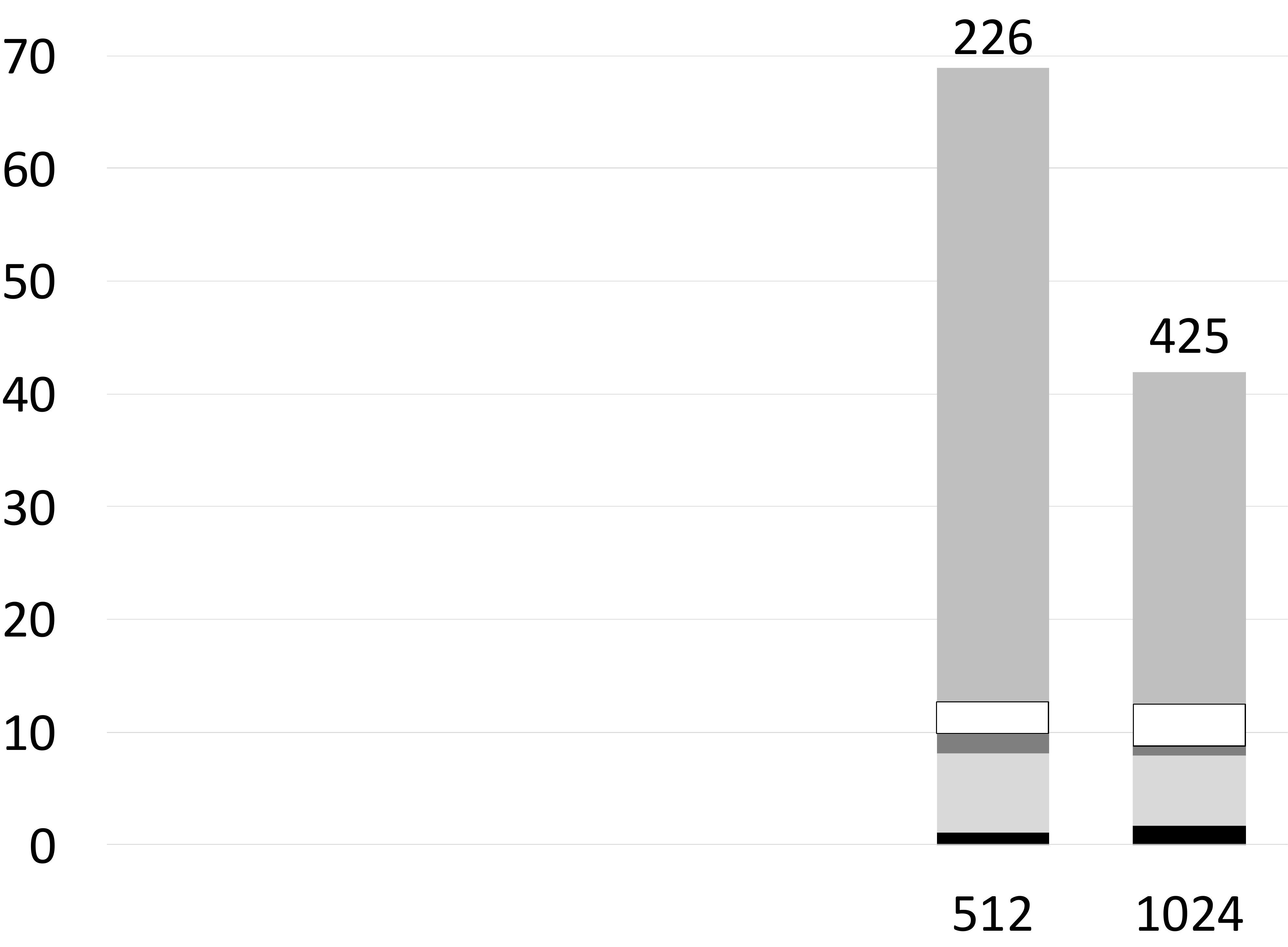}
        \caption{256 million elements}
        \label{fig:scal_256m}
    \end{subfigure}
    \caption{
      {\bf Cube benchmark.
      Strong scalability, parallel startup, various mesh sizes.}
      The parallel simulation startup phase is compared with 1000 steps of the Salvus time loop in terms of wall time.
      Approximate wall time for any number of timesteps can be obtained using simple proportionality.
      The number of timesteps in production simulations varies, but generally 30'000 or more are required.
      The particular stages are described in \cref{sec:startup}.
      {\bf X-axis}: number of Piz Daint nodes, each with 12 cores and 1 GPU.
      {\bf Y-axis:} wall time in seconds.
      {\bf Labels above bars:} the number of timesteps that take the same wall time as the startup phase.
      {\bf Missing bars:} out of memory failure during the time loop caused by the memory limit of the GPUs.
      {\bf Order of colors} in the bars is the same as in the legend.
      {\bf Note:} \subref{fig:scal_008m} is the same as \cref{fig:scal_016m_600s} but with a re-scaled time axis.
    }
    \label{fig:scal}
\end{figure}

\begin{figure}
    \begin{subfigure}{.495\columnwidth}
        \includegraphics[width=\linewidth]{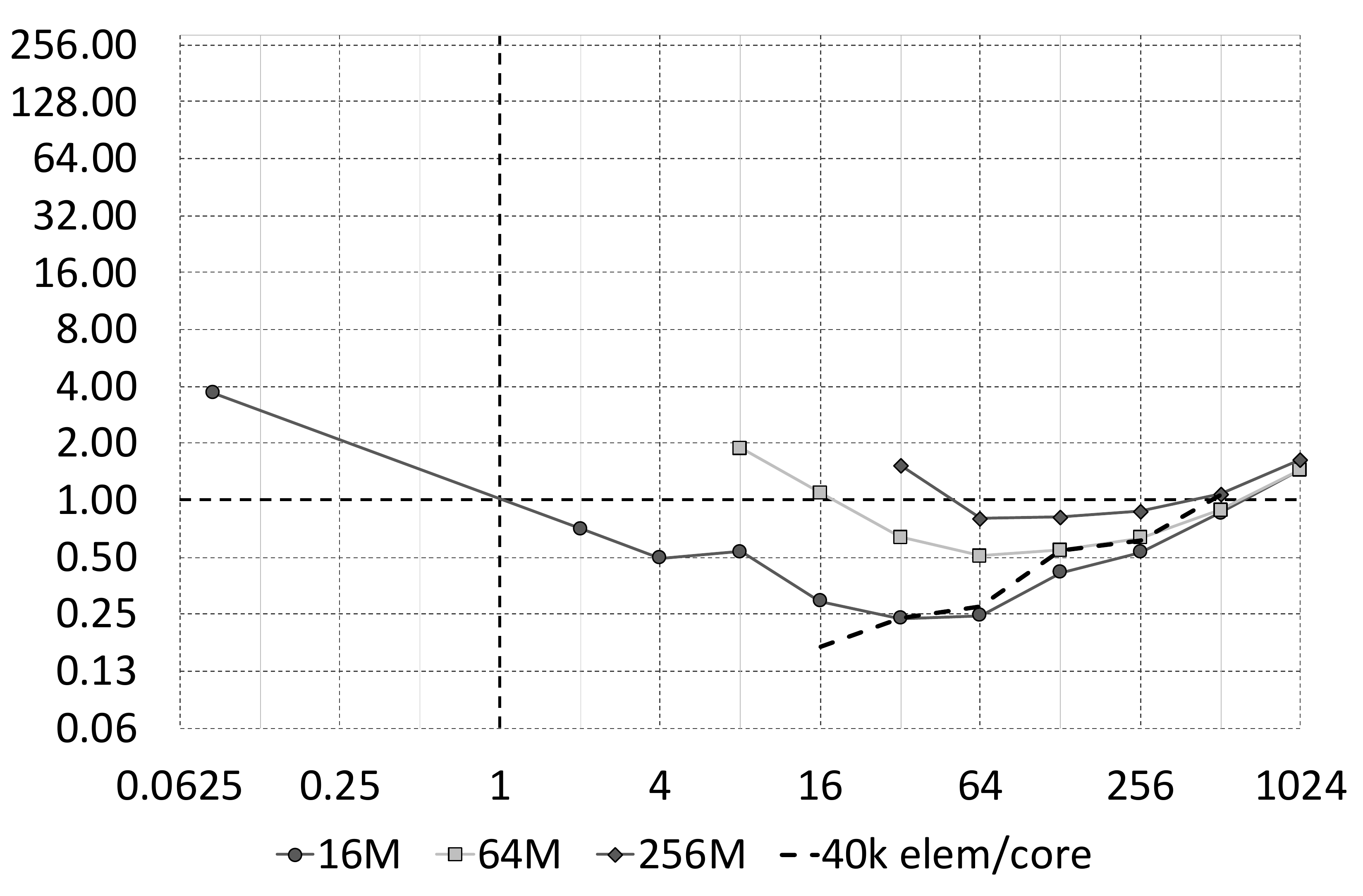}
        \caption{Raw data loading}
        \label{fig:scal_load}
    \end{subfigure}
    \hfill
    \begin{subfigure}{.495\columnwidth}
        \includegraphics[width=\linewidth]{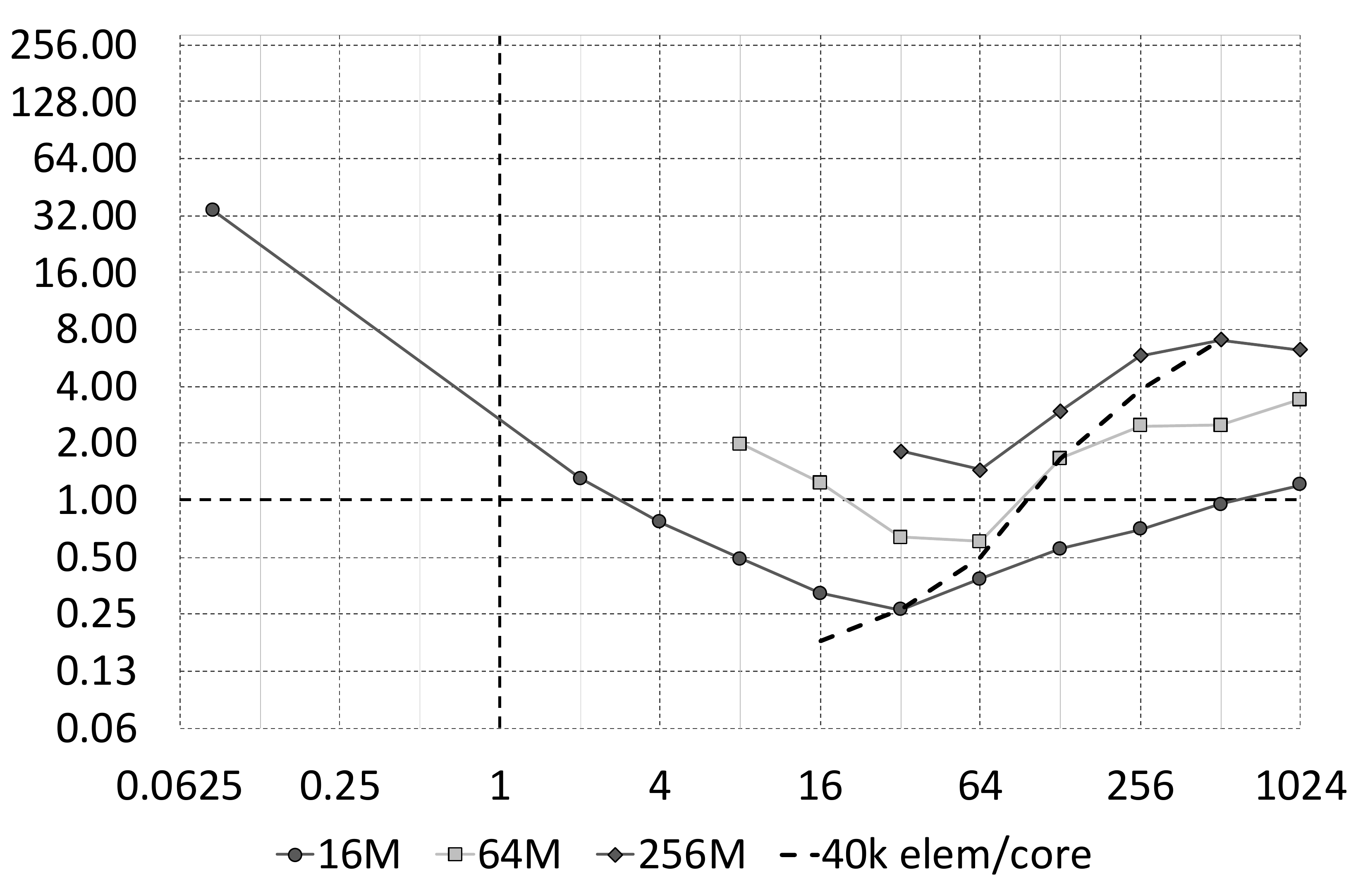}
        \caption{Distributed plex construction}
        \label{fig:scal_construct}
    \end{subfigure}

    \begin{subfigure}{.495\columnwidth}
        \includegraphics[width=\linewidth]{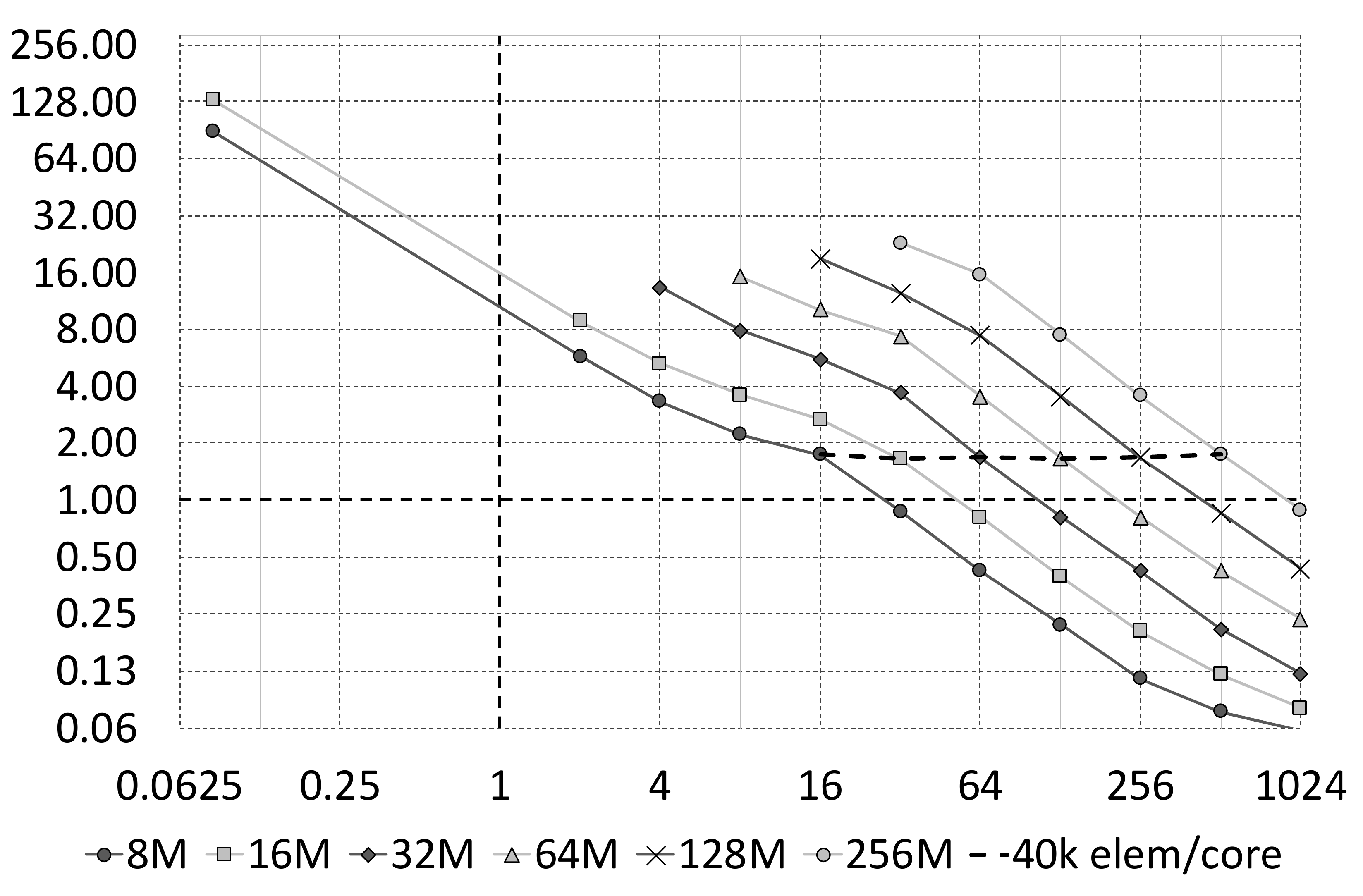}
        \caption{Topological interpolation}
        \label{fig:scal_interpolate}
    \end{subfigure}
    \hfill
    \begin{subfigure}{.495\columnwidth}
        \includegraphics[width=\linewidth]{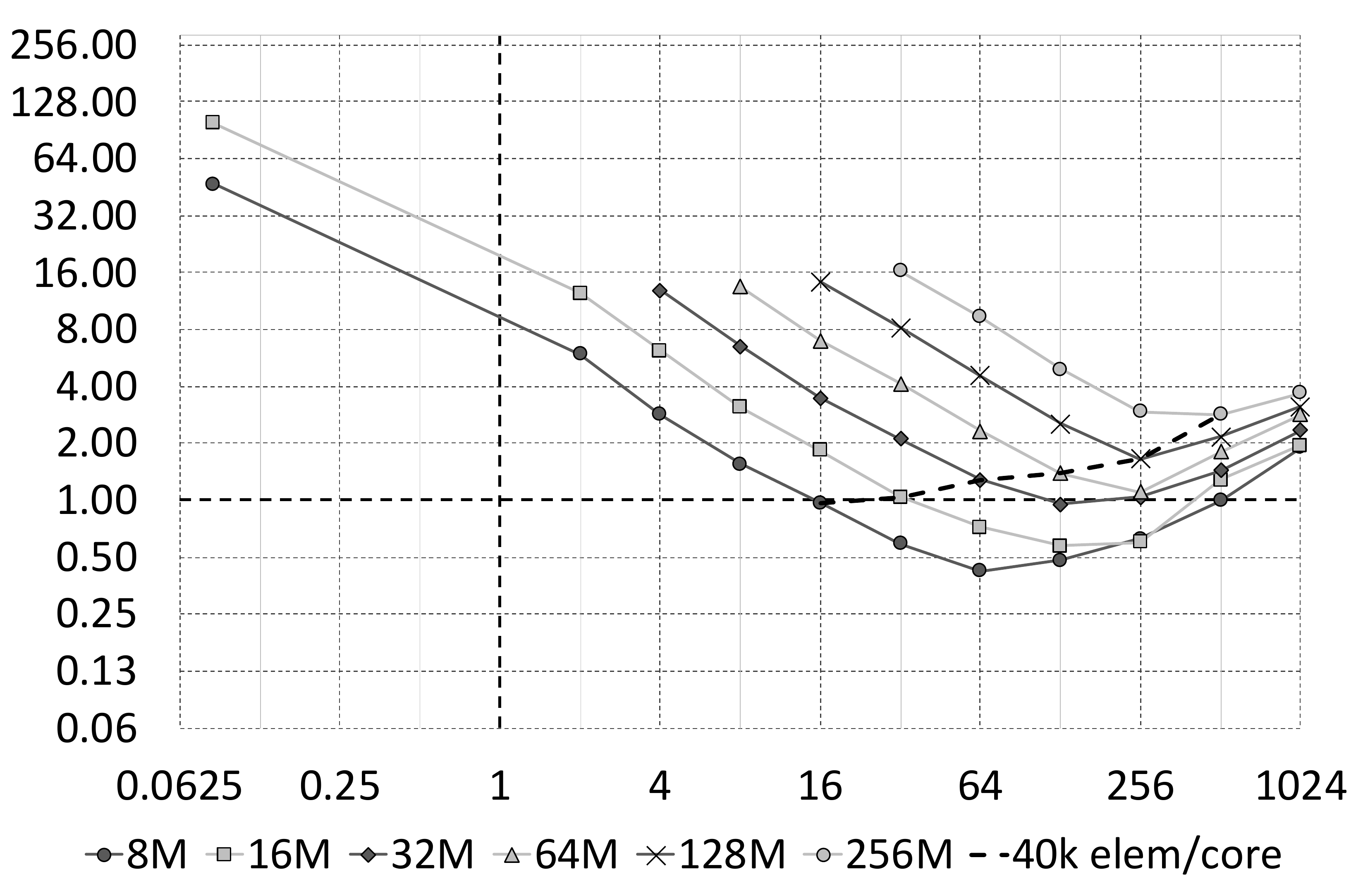}
        \caption{Redistribution}
        \label{fig:scal_distribute}
    \end{subfigure}

    \begin{subfigure}{.495\columnwidth}
        \includegraphics[width=\linewidth]{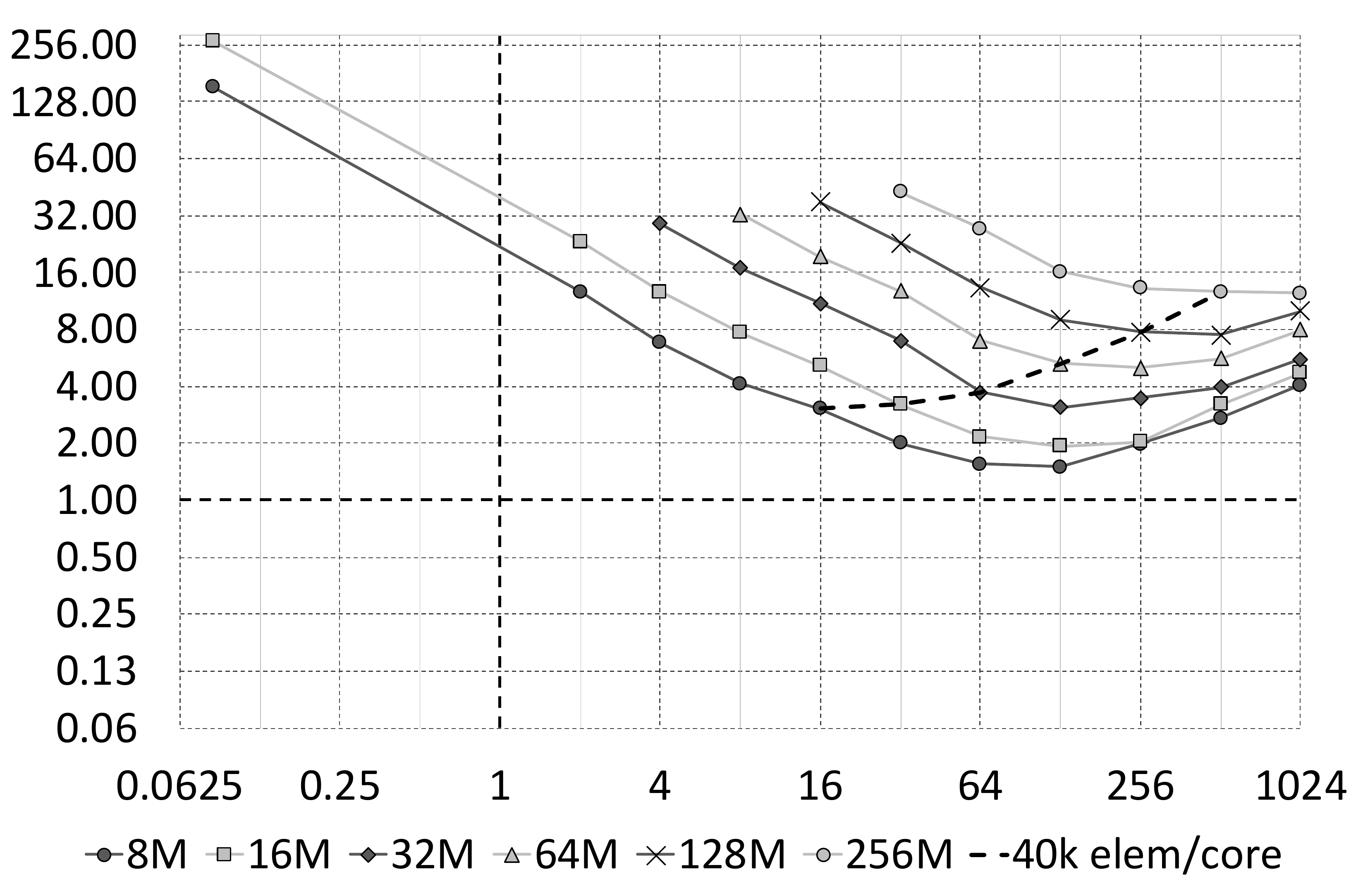}
        \caption{Total startup}
        \label{fig:scal_startup}
    \end{subfigure}
    \hfill
    \begin{subfigure}{.495\columnwidth}
        \includegraphics[width=\linewidth]{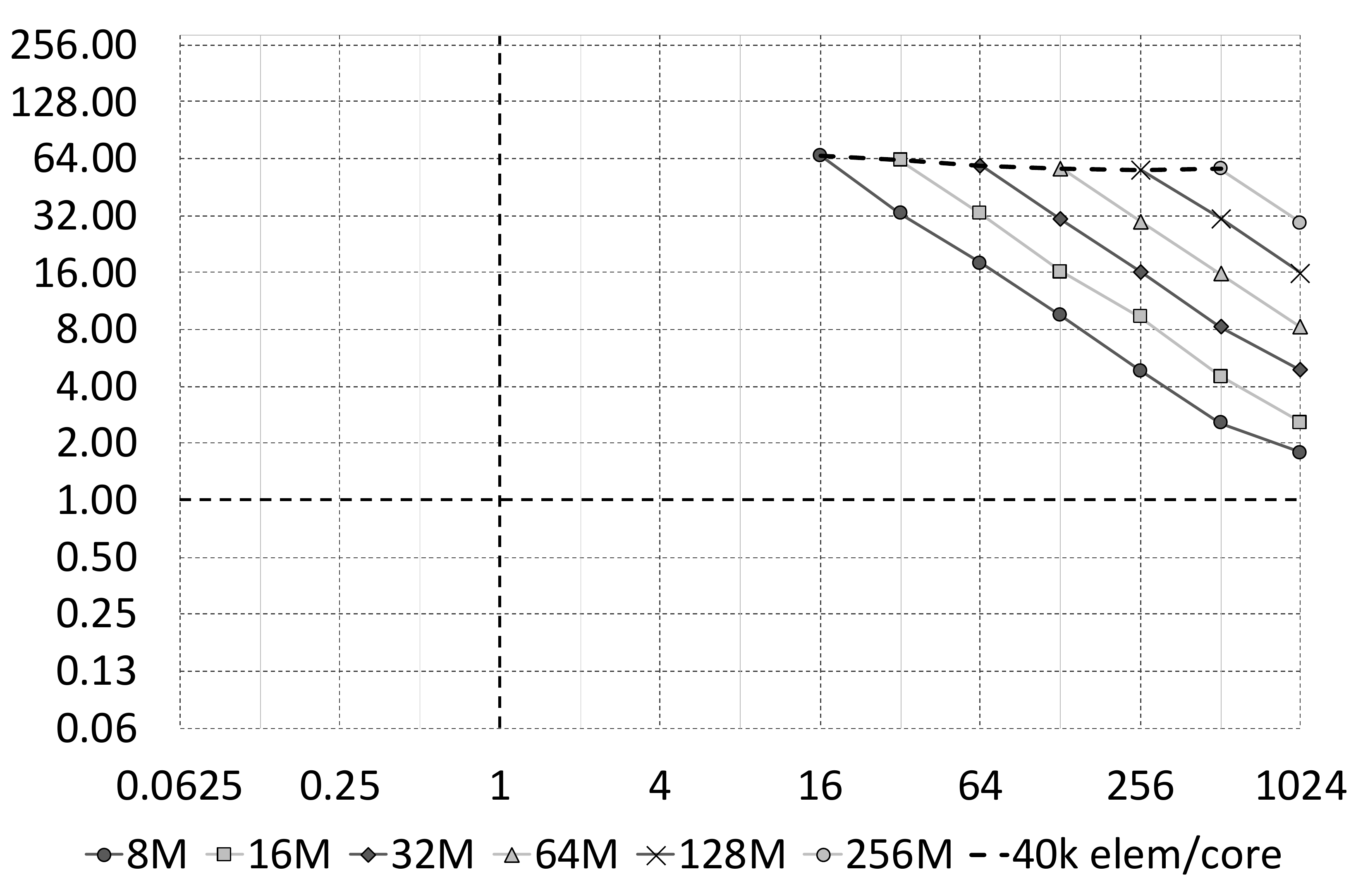}
        \caption{1000 Salvus timesteps}
        \label{fig:scal_timeloop}
    \end{subfigure}
    \caption{
      {\bf Cube benchmark. Strong and weak scalability per stage.}
      {\bf X-axis}: number of Piz Daint nodes (log2 scale), each with 12 cores and 1 GPU;
      $0.0625 = 1/12$ means a single core (serial) run.
      Note that a single node run was not possible due to the out-of-memory failure during the redistribution phase.
      {\bf Y-axis:} wall time in seconds (log2 scale).
      {\bf Solid lines} show the strong scalability for different mesh sizes. 
      {\bf Dashed line} shows the weak scalability for the mesh size of approximately 40'000 elements per core which is the upper bound for the Salvus timeloop imposed by limited GPU memory.
      {\bf Order of line styles} is the same in the plots and in the legends.
    }
    \label{fig:scal_per_stage}
\end{figure}

Graphs in \cref{fig:scal} are similar to \cref{fig:scal_016m_serial_vs_parallel} but show the only the parallel startup scalability for various mesh sizes, so that the displayed time can be limited to 70 seconds.
These graph gather all stages in a single graph and different mesh sizes are presented separately.
By contrast, graphs in \cref{fig:scal_per_stage} show strong and weak scalability for each stage separately, gathering all mesh sizes in a single graph.
We can see that the topological interpolation scales almost perfectly and becomes insignificant for high number of nodes even for very large meshes.
The other stages do not scale that well; however, their absolute wall times are rather small for the mesh sizes and node counts of interest. 
The scalability of the significant redistribution stage breaks at about 256 nodes and the mesh size no more dictates the wall time.
ParMetis was used with default settings; there might be some space for slight improvement by tuning its parameters but in general it is well known that current graph partitioners do not scale beyond 10'000 cores.
We can hardly do anything about it apart from perhaps testing alternative approaches such as space-filling curves.
There might be some space for improvement left for the distributed plex construction; nevertheless, from the stagnation between 256 and 1024 nodes for the largest mesh we conclude that such optimization is probably not worth the effort, at least for now.

\begin{table}[]
\centering
\small
\begin{tabular}{@{}rrrrrr@{}}
\toprule
\multicolumn{1}{l}{} &
  \multicolumn{2}{c}{topology (int64)} &
  \multicolumn{2}{c}{geometry (double)} &
  \multicolumn{1}{l}{} \\
\multicolumn{1}{l}{NEX} &
  \multicolumn{1}{l}{rows = NE = (NEX)$^3$} &
  \multicolumn{1}{l}{columns} &
  \multicolumn{1}{l}{rows} &
  \multicolumn{1}{l}{columns} &
  \multicolumn{1}{l}{file size (GB)} \\
\midrule
200 & 8'000'000   & 8 & 8'120'601   & 3 & 0.66  \\
252 & 16'003'008  & 8 & 16'194'277  & 3 & 1.32  \\
318 & 32'157'432  & 8 & 32'461'759  & 3 & 2.64  \\
400 & 64'000'000  & 8 & 64'481'201  & 3 & 5.26  \\
504 & 128'024'064 & 8 & 128'787'625 & 3 & 10.51 \\
635 & 256'047'875 & 8 & 257'259'456 & 3 & 21.01 \\ \bottomrule
\end{tabular}
\caption{Cube benchmark data files: number of elements in x-direction (NEX); total number of elements (NE); topology (connectivity) and geometry (vertices) dataset sizes; file sizes. Both topology (integer numbers) and geometry (real numbers) are stored with 64-bit precision.}
\label{tab:file_sizes}
\end{table}

\section{Application: seismicity on Mars}\label{sec:results}
In late 2018, the NASA InSight mission \cite{Banerdt2020} placed a highly sensitive seismometer \cite{Lognonne2019} on Mars' surface and recorded the first seismic signals ever observed on Mars \cite{Giardini2020}.
The observation of seismic waves is a crucial source of information to investigate the interior structure and composition of Mars.
However, as the data shows significant differences to seismic data from both Moon and Earth,
numerical simulations of seismic wave propagation on Mars that account for topography as well as 3D scattering due to lateral variations of the material parameters are key to assist the interpretation of the observational data.

Full-waveform simulations are essential to constrain the planet's structure using data in the frequency band recorded by the probe.
The seismic response to marsquakes or asteroid impacts is governed by a coupled system of the elastic/acoustic wave equation,
which models seismic waves propagating through Mars' mantle and the liquid core, respectively.
This can be simulated efficiently using the spectral-element method (SEM).

\begin{figure}
  \begin{subfigure}{\columnwidth}
    \centering
    \includegraphics[width=.5\linewidth]{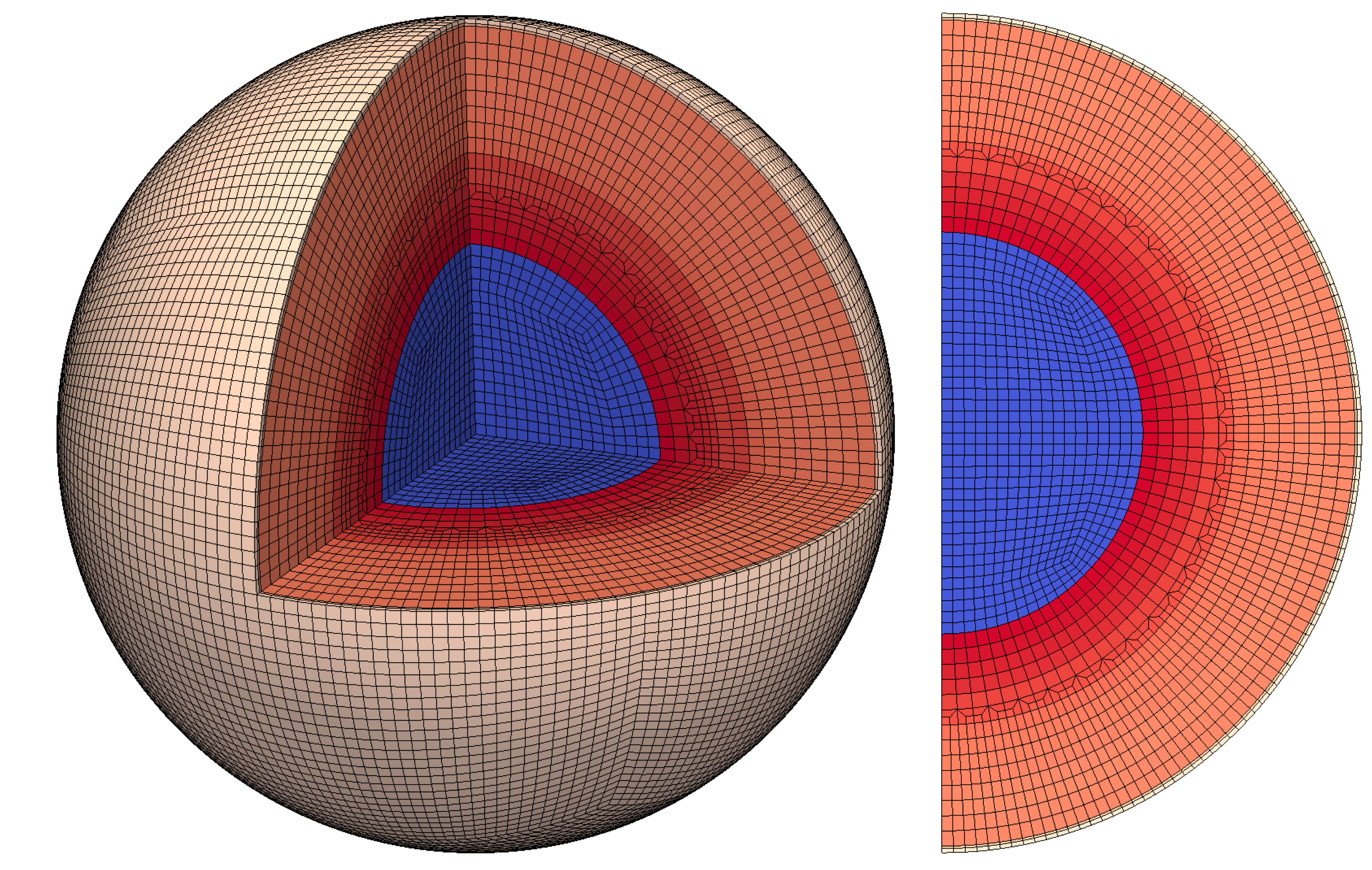}
    \caption{cubed sphere with fluid core (blue) and solid mantle (red)}
  \end{subfigure}

  \begin{subfigure}{\columnwidth}
    \centering
    \includegraphics[width=.5\linewidth]{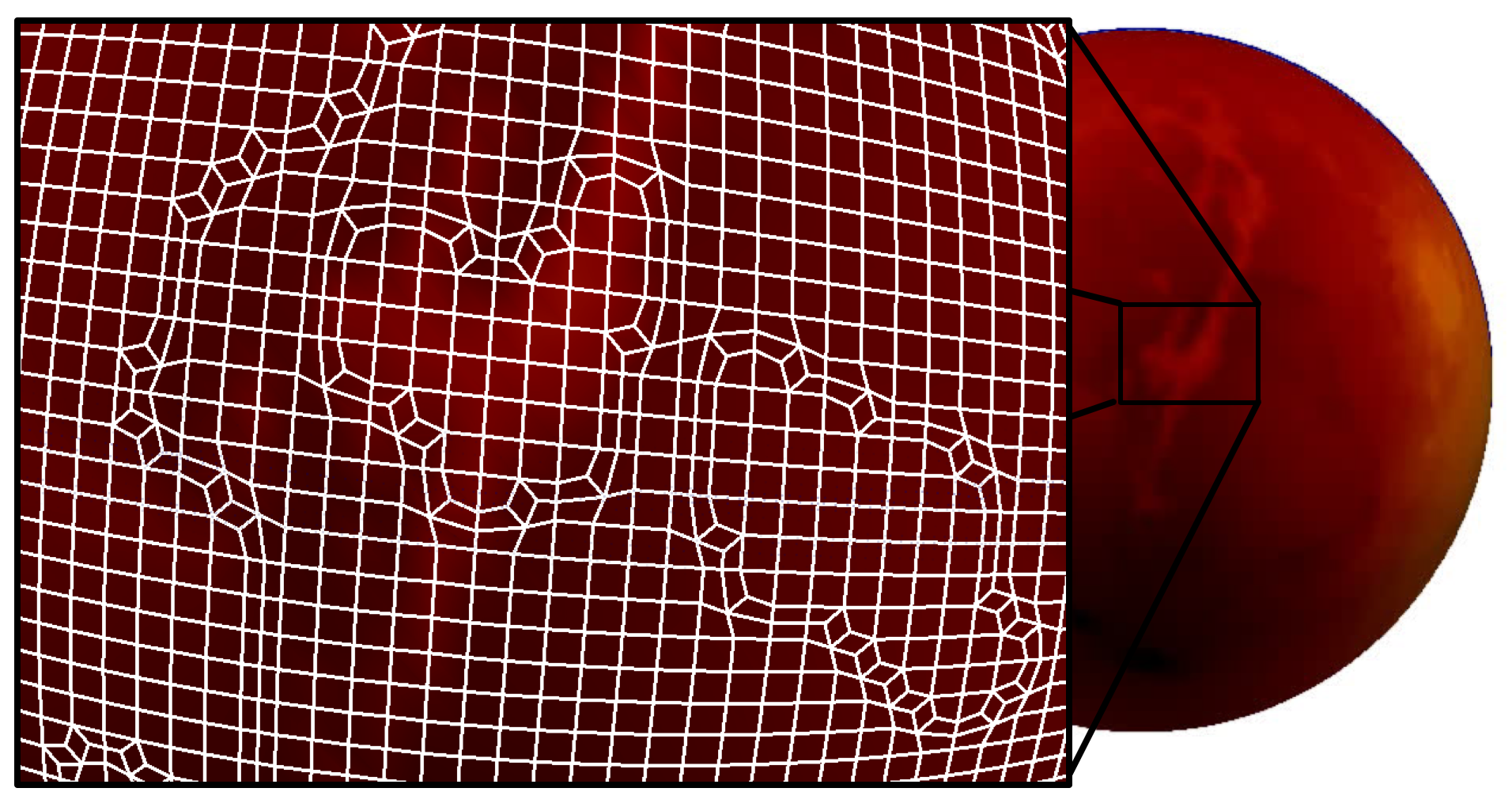}
    \caption{modeling 3D topography on the surface}
  \end{subfigure}

  \begin{subfigure}{\columnwidth}
    \centering
    \includegraphics[width=.5\linewidth]{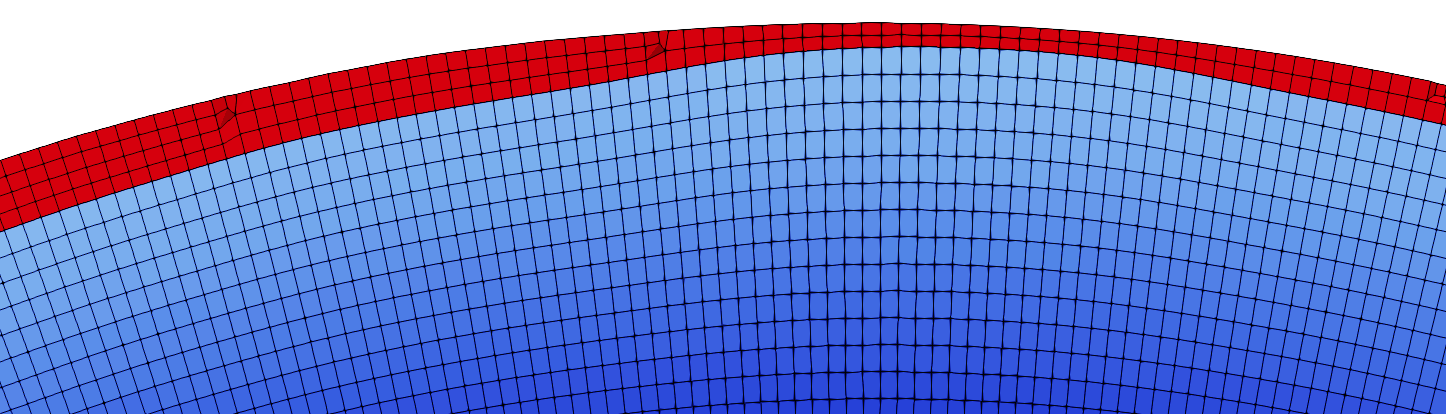}
    \caption{crustal thickness variations (red)}
  \end{subfigure}
  
  \begin{subfigure}{\columnwidth}
    \centering
    \includegraphics[width=0.35\linewidth]{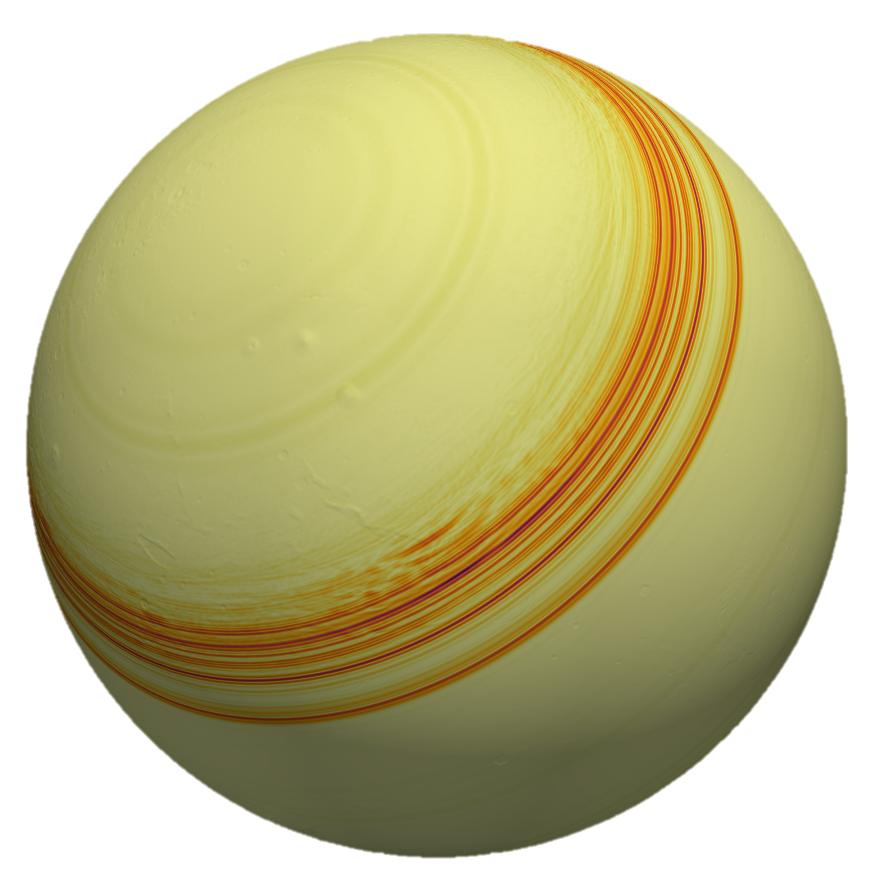}
    \caption{snapshot of the seismic wavefield on the surface (result of the simulation)}
    \label{fig:mars_snap}
  \end{subfigure}

  \caption{Anisotropic mesh refinements to accurately model the structure of Mars on a conforming hexahedral mesh.
    The Mars texture map is based on NASA elevation and imagery data.}
  \label{fig:mars_mesh}
\end{figure}

For the computation of the seismic response of Mars, we rely on Salvus' implementation of the SEM (\cref{sec:salvus}).
Salvus' internal mesher uses custom algorithms to generate fully unstructured conforming 3D hexahedral meshes \cite{van_driel_accelerating_2020},
efficiently representing topography and the extreme crustal thickness variations of Mars ($\sim$5--120 km), see \cref{fig:mars_mesh}.
The solver then represents these meshes in memory using DMPlex (\cref{sec:dmplex}).

The CFL condition for Salvus' explicit second-order Newmark time-stepping scheme, coupled with a required minimum number of points-per-wavelength in each dimension, results in the computational complexity of a simulation scaling with frequency to the power of 4.
When using 4-th order spectral elements, which is common for planetary-scale wave propagation, more than 6 grid points per shortest wave-length are needed to accurately resolve seismic waves \cite{Fichtner_book}.
As the quakes are small in magnitude and the noise level increases at low frequencies, large-scale simulations are required to reach the parameter regime of the observations,
and the required number of spectral elements can easily reach hundreds of millions.

Such mesh sizes necessitate the parallel simulation startup presented in \cref{sec:startup}.
Prior to these developments, the largest possible mesh size was limited by the available memory of a single Piz Daint node to approximately 16 million elements.
Moreover, loading a mesh of such size took more than four minutes.
With the parallel startup in hand, these limitations vanish.

\Cref{fig:mars_snap} shows a snapshot of the surface displacement resulting from a simulation of a hypothetical quake on Mars.
Here, the discretized coupled elastic wave equation has approximately 124 million 4-th order spectral elements, and we compute 100'455 timesteps representing a simulated time of 30 minutes. 
Each element has 125 spatial DOFs, each hosting 9 dynamic field components (vector displacement, velocity, and acceleration). 
These parameters lead to an unprecedented resolved period of 3.2 s.
Using again all 12 cores per Piz Daint node as well as the attached Tesla P100 GPU,
this simulation took approximately 2.4 hours of wall time on 256 Piz Daint nodes.
From this total wall time, raw data loading took 0.9 s, distributed plex construction 4.5 s, topological interpolation 3.1 s and redistribution 2.3 s, i.e., the whole startup phase took less than 11 s.

\section{Conclusions}
We presented algorithmic strategies for handling unstructured meshes for high-order
finite-element simulations on complex domains. In particular, we demonstrated new
capabilities for parallel mesh reading and topological interpolation in PETSc DMPlex,
which enables a fully parallel workflow starting from the initial data file reading.
This is beneficial not only for direct users of DMPlex but also users of software libraries and packages employing DMPlex,
such as Firedrake \cite{rathgeber_firedrake_2016}, Salvus \cite{Afanasiev_2019}, or HPDDM \cite{hpddm}.

This work in a sense follows up~\cite{lange_flexible_2015}
and addresses the main task stated in their Future Work:
{\em
``Most crucially perhaps is the development of a fully parallel mesh input reader in
PETSc in order to overcome the remaining sequential bottleneck during model initialisation.''
}
Moreover, that paper mentions the ``HDF5-based XDMF output format'' but it has become an {\em input} format as well
within this work. Hence, HDF5/XDMF has become the first widely used mesh format supported by PETSc which is both readable and writable in parallel.

The implementation is agnostic to the type of finite elements in the mesh and completely
decoupled from the governing equations, and is thus applicable in many scientific disciplines.
In particular, our solution overcomes bottlenecks in numerical modeling of seismic wave propagation
on Mars and shows excellent parallel scalability in large-scale simulations on more
than 12'000 cores.

\section*{Acknowledgments}
All presented PETSc developments have been made publicly available in PETSc since its release 3.13.

We gratefully acknowledge support from
the Swiss National Supercomputing Centre (CSCS) under projects s922 and s961;
the Platform for Advanced Scientific Computing (PASC) under the project ``Salvus'';
the Swiss National Science Foundation (SNF) BRIDGE fellowship program under the grant agreement No. 175322;
the European Research Council (ERC) from the EU's Horizon 2020 programme under grant agreement No. 714069;
the EU Centre of Excellence ChEESE under grant agreement No. 823844;
and the ETH Zurich Postdoctoral Fellowship Program which received funding from the EU’s Seventh Framework Programme under the grant agreement No. 608881.

\bibliographystyle{siamplain}
\bibliography{hapla-mesh-io}
\end{document}